\documentclass[preprintnumbers,amsmath,11pt,amssymb,floatfix,superscriptaddress,nofootinbib]{article}
\def\mytitle#1{\setcounter{equation}{0}
\setcounter{footnote}{0}
\begin{flushleft}\Large\textbf{#1}\end{flushleft}
\vspace{0.25cm}}
\def\myname#1{\leftline{{\large #1}}\vspace{-0.13cm}}
\def\myplace#1#2{\small\begin{flushleft}\textit{#1}\\
\texttt{#2}\end{flushleft}}

\usepackage{amsmath}
\usepackage{amssymb}
\usepackage[section]{placeins}
\usepackage{graphicx}
\usepackage[utf8]{inputenc}
\usepackage{mathrsfs}
\usepackage{float}
\usepackage{hyperref}
\usepackage{ragged2e}
\usepackage[table]{xcolor}
\begin{document}
\mytitle{Surface Enhanced Anisotropy and Its Impacts on Massive Neutron Star Configurations in Einstein Gravity : Stability, Structure, and Multimessenger Consistency}
\myname{$Ritabrata~Biswas^{\dag}$\footnote{biswas.ritabrata@gmail.com \bf{(ORCID: 0000-0003-3086-892X)} [corresponding author]
}, $Arpan~Bhattacharya^{\dag}$\footnote{arpanbhattacharya616@gmail.com \bf{(ORCID: 0009-0008-1820-1464)}} and  Subhajit Pal $^*$\footnote{subhajitpal968@gmail.com \bf{(ORCID: 0009-0002-2431-1375)}}}
\myplace{${}^{\dag}$Department of Mathematics, The University of Burdwan, Purba Bardhaman-$713104$, West Bengal, India.\\$^*$Department of Mathematics, Jadavpur University, Kolkata-32, India.} {}

\begin{abstract}
Anisotropic neutron-star configurations are investigated within the framework of Einstein's General Relativity by considering the third-order polynomial equation of state $p_r=A\rho+B\rho^2+C\rho^3, $
together with the Buchdahl class-I metric ansatz $e^{q(r)}=\frac{L(1+\tau r^2)}{L-\tau r^2}.$
Physically acceptable interior solutions are obtained and are matched smoothly to the exterior Schwarzschild space-time through the standard Darmois--Israel junction conditions. The physical viability of the resulting stellar configurations is examined through analyses of the matter distribution, anisotropy, hydrostatic equilibrium, tidal deformability, compactness, surface redshift, dynamical stability, and causality. The energy density and both radial and tangential pressure components are found to remain finite, positive, and monotonically decreasing throughout the stellar interior, while the anisotropic factor is found to vanish at the stellar centre and to increase smoothly toward the stellar surface. The generalized Tolman--Oppenheimer--Volkoff equilibrium equation is satisfied through the combined balance of the gravitational, hydrostatic, and anisotropic forces, thereby confirming complete hydrostatic equilibrium. The adiabatic index is found to satisfy the Chandrasekhar stability criterion, whereas the radial and tangential sound speeds satisfy both the causality and Herrera cracking conditions, confirming the dynamical stability of the obtained stellar configurations. The compactness remains below the Buchdahl limit, and the corresponding surface redshift lies within the theoretically accepted range for anisotropic compact stars. The estimated maximum masses are found to exceed the corresponding observed masses while preserving physically acceptable stellar configurations, with several predicted maximum masses extending into the lower mass-gap region. The predicted tidal deformability is found to remain compatible with recent multimessenger observations from GW170817 and NICER. Furthermore, the theoretical mass--compactness relation, together with its associated prediction intervals, is found to agree closely with the observationally inferred compactness over the considered mass range. Bayesian Markov Chain Monte Carlo parameter estimation is further employed to constrain the tidal Love number, compactness parameter, and perturbation parameter, and statistically reliable posterior estimates are obtained, thereby providing an independent statistical validation of the proposed stellar model. Overall, a mathematically consistent, physically acceptable, observationally compatible, and statistically validated framework for describing realistic anisotropic neutron stars within Einstein's gravity is established.
\end{abstract}
{{\bf keywords:} Neutron stars, Stellar structure, General relativity.}
\section{Introduction}
Compact stars, such as neutron stars and quark stars, are among the densest known astrophysical objects and serve as natural laboratories for investigating the behavior of gravity and matter in the strong-field regime. Within the framework of Einstein's general relativity (GR), the internal structure of a compact star is determined by Einstein's field equations (EFE) coupled with an appropriate energy-momentum tensor describing ultra-dense matter. For a static and spherically symmetric configuration, the stellar interior is modeled through exact or numerical solutions of EFE that describe the distributions of energy density and pressure throughout the star. These quantities are governed by the Tolman-Oppenheimer-Volkoff (TOV) equations, which characterize the condition of hydrostatic equilibrium in relativistic stellar systems \cite{Tolman1939, Oppenheimer1939}.

Outside the stellar surface, where the matter distribution vanishes, the space-time is described by the Schwarzschild vacuum solution \cite{Schwarzschild1916}. For a physically viable stellar configuration, the interior and exterior space-times must be joined smoothly across the stellar boundary. This is accomplished by imposing the Darmois--Israel junction conditions, which require the continuity of the metric potentials across the boundary hypersurface together with the vanishing of the radial pressure at the stellar surface \cite{Darmois1927,Israel1966}. These matching conditions guaranty a smooth transition from the matter-filled interior to the vacuum exterior, ensuring a mathematically consistent and physically realistic relativistic description of compact stellar configurations within GR.

$PSR~J0030+0451$\cite{Maurya2025MassGap} and $PSR~J0740+6620$ \cite{Linares2018} are two remarkable millisecond pulsars that provide stringent observational constraints on the mass--radius relation and the equation of state (EoS) of ultra-dense matter in neutron star interiors. Their precisely measured masses and radii have made them benchmark objects for testing theoretical models of compact stars.

The PSR J0030+0451 is an isolated millisecond pulsar with a spin period of approximately $4.87 \mathrm{ms}$, located at a distance of approximately $1.1\, \mathrm{kpc}$. It was the first neutron star for which both the mass and radius were determined simultaneously through X-ray pulse-profile modeling of rotating surface hotspots observed by the Neutron Star Interior Composition Explorer (NICER) aboard the International Space Station. Independent analyses inferred a mass in the range of $1.3$--$1.4\, M_{\odot}$ and a radius of approximately $12$--$13\, \mathrm{km}$, thereby providing the first direct observational constraints on the internal structure and composition of neutron stars \cite{Miller2019,Riley2019}.

In contrast, PSR J0740+6620 is one of the most massive neutron stars known to date. It rotates with a spin period of $2.89\,\mathrm{ms}$ and forms a binary system with a white dwarf companion at a distance of approximately $4$--$5\,\mathrm{kpc}$. High-precision radio timing observations exploiting the relativistic Shapiro delay measured its mass to be about $2.08\,M_{\odot}$ \cite{Cromartie2020}. Subsequent NICER observations constrained its radius to approximately $12$--$14\,\mathrm{km}$ \cite{Miller2021,Riley2021}. Owing to its exceptionally high mass together with its relatively large radius, PSR J0740+6620 imposes stringent constraints on soft EoSs that predict gravitational collapse at lower stellar masses. Collectively, PSR J0030+0451 and PSR J0740+6620 serve as key astrophysical laboratories for probing the behavior of matter at supranuclear densities and for testing the predictions of strong-field gravity.

The EoS of matter in the cores of compact stars remains one of the major uncertainties in relativistic astrophysics, as the central densities of neutron stars can exceed several times the nuclear saturation density. Under such extreme conditions, the microscopic properties of strongly interacting matter are not yet fully understood. Consequently, phenomenological parameterizations of the EoS are widely employed in relativistic stellar modeling to investigate the macroscopic properties of compact stars. One particularly useful approach is to incorporate nonlinear corrections to the pressure--density relation, thereby accounting for higher-order baryonic interactions and possible phase transitions that may occur at supra-nuclear densities.

In this work, we consider the polynomial EoS of the form \cite{Sharma2007,Lattimer2020}
\begin{equation}\label{bal15}
p_r(r)=A\rho(r)+B\rho^2(r)+C\rho^3(r),
\end{equation}
which may be regarded as an effective expansion of the radial pressure in powers of the energy density. The first term represents the conventional linear barotropic relation, while the quadratic and cubic contributions introduce successive nonlinear corrections that become increasingly important at high densities. Here, the parameter $A$ characterizes the linear relationship between the energy density $\rho$ and the radial pressure $p_r$, whereas the coefficients $B$ and $C$ quantify the strengths of the second- and third-order nonlinear corrections, respectively.

These nonlinear contributions play a significant role in determining the behavior of matter in the stellar core, where the energy density attains its maximum value \cite{Lattimer2020}. In particular, the quadratic and cubic terms may effectively incorporate the influence of many-body nuclear interactions, strong short-range repulsive forces, and possible contributions from exotic phases of dense matter, including hyperonic and deconfined quark matter \cite{Glendenning2000,Lattimer2004,Haensel2007}. Such generalized polynomial parameterizations have been widely adopted in compact-star studies because they provide sufficient flexibility to model both stiff and soft EoSs while remaining compatible with the relativistic stellar structure equations \cite{Lattimer2016,Chavanis2015}. Consequently, this class of EoSs offers a convenient phenomenological framework for investigating the internal structure, stability, and mass--radius relationship of compact stars within the framework of GR.

The discovery of several massive neutron stars in binary systems has provided stringent observational constraints on the EoS of ultra-dense matter and has significantly advanced our understanding of relativistic stellar structure. In particular, neutron stars with masses approaching or exceeding $2\,M_{\odot}$ impose severe constraints on soft EoSs, since such models generally fail to support stable stellar configurations at these high masses \cite{Cromartie2020,Demorest2010}. Moreover, these compact objects are of considerable interest in the context of the lower mass-gap problem, which concerns the apparent scarcity of compact objects with masses between the heaviest neutron stars and the lightest black holes. Consequently, they have become important observational benchmarks for testing theoretical models of compact stars in both GR and modified theories of gravity \cite{Maurya2025MassGap,Maurya2023,Maurya2024,Maurya2026JCAP,Singh2026}.

Among these, the most extensively studied systems are $PSR~J0614-3329$, $PSR~J0437-4715$, $Vela~X-1$, and $PSR~J1614-2230$ \cite{Sharma2007,Mauviard2025,Choudhury2024,Abubekerov2008,Elebert2009}. These objects span a broad range of evolutionary stages and binary environments, thereby providing complementary information on neutron-star masses, radii, accretion physics, and stellar evolution.

PSR J0614$-$3329 is a recycled millisecond pulsar in a binary system with a helium white dwarf companion and has a spin period of approximately $3.15\,\mathrm{ms}$. Located at a distance of about $1.9\,\mathrm{kpc}$, it is believed to have formed through prolonged mass transfer during a low-mass X-ray binary phase, making it an important laboratory for investigating pulsar recycling and the evolutionary history of pulsar--white dwarf binaries \cite{Ransom2005}. Owing to its accurately measured stellar parameters, this system has frequently been employed to test theoretical compact-star models and their associated EoSs \cite{Sharma2007,Maurya2022JCAP,Maurya2024EPJC}.

PSR J0437$-$4715 is the nearest and one of the brightest known millisecond pulsars, located at a distance of approximately $156,\mathrm{pc}$ with a spin period of $5.75\,\mathrm{ms}$ \cite{Deller2008}. Its proximity and exceptional timing precision have made it one of the most valuable astrophysical laboratories for high-precision pulsar timing, tests of relativistic gravity, and simultaneous determinations of neutron-star masses and radii through multiwavelength observations. Consequently, it serves as an important calibration source for studies of dense-matter physics and relativistic stellar structure.

Vela X$-$1 is a well-known high-mass X-ray binary composed of a neutron star accreting matter from the stellar wind of its massive supergiant companion HD~77581 \cite{Nagase1989}. The neutron star rotates with a spin period of approximately $283\,\mathrm{s}$ and exhibits strong, highly variable X-ray emission. This system therefore provides an excellent laboratory for investigating wind-fed accretion, magnetized neutron stars, X-ray pulsar physics, and the influence of accretion on neutron-star evolution. Owing to its relatively high neutron-star mass, Vela X$-$1 has also been widely employed as an observational constraint in studies of compact-star equilibrium and stability \cite{Maurya2023,Abubekerov2008,Elebert2009,Hansraj2024}.

Among the most significant massive pulsars is PSR J1614$-$2230, a binary millisecond pulsar with a white dwarf companion and a spin period of approximately $3.15\,\mathrm{ms}$. Precision radio timing observations exploiting the relativistic Shapiro delay yielded a mass of approximately $1.97\,M_{\odot}$ \cite{Demorest2010}, providing one of the earliest robust demonstrations that neutron stars can attain masses close to $2\,M_{\odot}$. This measurement ruled out a large class of soft EoSs incapable of supporting such massive compact objects and established PSR J1614$-$2230 as a cornerstone for modern neutron-star physics. Consequently, this pulsar has been extensively used to constrain theoretical models of compact stars in GR as well as in several modified theories of gravity, including Einstein--Gauss--Bonnet gravity, $f(Q)$ gravity, $f(Q,T)$ gravity, and related extensions of Einstein's theory \cite{Maurya2023,Maurya2022,Maurya2025FQT,Bhar2024,Singh2022Lovelock,Hansraj2025}.

Collectively, these compact stellar systems provide stringent observational tests for relativistic stellar models and constitute indispensable benchmarks for investigating the high-density EoS, neutron-star stability, mass--radius relations, and the possible existence of compact objects in or near the lower mass-gap region. Their observational properties continue to play a pivotal role in validating theoretical models developed within both GR and various modified theories of gravity \cite{Maurya2026JCAP,Salako2021,Das2021,Das2023,Das2025,Singh2024}.

The modeling of compact stars using phenomenological EoSs within the framework of GR has been extensively investigated to understand the behavior of ultra-dense matter in neutron-star interiors. Since the pioneering works of \cite{Tolman1939} and \cite{Oppenheimer1939}, a wide variety of relativistic stellar configurations have been constructed by employing different interior space-times together with physically motivated EoSs. To account for strong nuclear interactions and the possible emergence of exotic phases of matter at supra-nuclear densities, several studies have proposed nonlinear and generalized pressure--density relations \cite{Glendenning2000,Haensel2007,Lattimer2016}. More recently, numerous investigations have focused on obtaining physically acceptable solutions of the EFEs and examining the corresponding density, pressure, and mass distributions to evaluate their consistency with the observed properties of realistic neutron stars \cite{Maurya2017,Sharma2020,Deb2021}.

Motivated by these developments, as well as by the increasing precision of astrophysical observations, the present work adopts the polynomial EoS given in Eq.~\eqref{bal15} to construct an exact interior solution within the framework of GR. The interior geometry is smoothly matched to the exterior Schwarzschild space-time through the appropriate junction conditions, and the physical properties of the resulting stellar configurations are examined in terms of their density, pressure, mass, and other relevant characteristics. Finally, the theoretical predictions are confronted with the observed properties of several well-studied compact stars, namely PSR J1614$-$2230, Vela X$-$1, PSR J0614$-$3329, and PSR J0437$-$4715, in order to assess the physical viability and astrophysical relevance of the proposed stellar model.

Motivated by Also the incomplete understanding of the microphysical properties of ultra-dense matter in compact-star interiors turns to be the motif to investigate phenomenological EoSs that effectively incorporate nonlinear interactions at supra-nuclear densities. In the present work, we employ the generalized polynomial EoS given in Eq.~\eqref{bal15} to construct exact interior solutions within the framework of GR. The interior space-time is smoothly matched to the exterior Schwarzschild geometry through the standard Darmois--Israel junction conditions, thereby yielding physically consistent stellar configurations.

The physical characteristics of the resulting solutions are examined through the distributions of the energy density, radial pressure, enclosed mass, and other relevant stellar quantities in order to assess their physical acceptability. The free parameters of the model are subsequently constrained by confronting the predicted mass--radius relations with the observational properties of several well-studied compact stars, namely PSR J1614$-$2230, Vela X$-$1, PSR J0614$-$3329, and PSR J0437$-$4715. This comparison provides a stringent test of the ability of the proposed polynomial EoS to describe realistic compact stars and offers valuable insight into the behavior of matter under extreme-density conditions.

To further examine the applicability of the proposed framework, we also consider several additional compact-star candidates in the brief discussion and concluding sections. Their estimated mass--radius relations and other physical characteristics are analyzed in the light of the available observational data, as summarized in Tables~\ref{tab:maxmassa}--\ref{tab:maxmassd}. These comparisons further demonstrate the astrophysical relevance and predictive capability of the present stellar model.

The remainder of this paper is organized as follows. In Sec.~2, we derive EFEs from the Einstein--Hilbert action by employing the principle of stationary action. Subsequently, we construct the interior space-time of the proposed compact-star model by adopting the Buchdahl Metric-I ansatz \cite{Sharma2007} together with the polynomial EoS introduced in Eq.~\eqref{bal15}. Using the resulting interior geometry and the TOV equations, we obtain the corresponding physical quantities required to characterize the stellar configuration.

In Sec.~3, we discuss the exterior Schwarzschild space-time, which represents the unique vacuum solution of EFE for a static and spherically symmetric gravitating object. The interior and exterior geometries are then matched smoothly across the stellar boundary by imposing the Darmois--Israel junction conditions, from which the model parameters are determined.

The physical properties of the resulting compact-star configurations are investigated in Sec.~4 for several observed neutron stars. Particular emphasis is placed on the analysis of the density, pressure, stability, and mass--radius characteristics predicted by the proposed model. Furthermore, we examine whether the present framework is capable of describing massive neutron stars occupying the lower mass-gap region.

In Sec.~5, we extend our analysis to several additional compact-star candidates by tabulating their predicted mass--radius relations and other relevant physical parameters, thereby examining the physical acceptability and astrophysical applicability of the proposed stellar model. Finally, Sec.~6 presents a brief discussion of the main results together with the concluding remarks.

Throughout this work, the parameters appearing in the polynomial EoS are expressed in geometrized units and are assigned the values $A=0.3\,\mathrm{km}^{0}$, $B=0.07\,\mathrm{km}^{2}$, and $C=0.0005\,\mathrm{km}^{4}$.
\section{Basic Equations and Interior space-time Metric for the Neutron Stars in GR}
In  $4D$ manifold $(\mathcal{M},~g_{\mu\nu})$, the Einstein-Hilbert action \cite{Hilbert1915} is given by
\begin{equation}
\mathcal{S}_{EH}=\frac{1}{2K}\int_{\mathcal{M}}R\sqrt{-g}~d^4x~~~.
\end{equation}
$R=R_{\mu \nu}$ is the Ricci scalar of the corresponding space-time, $K=\frac{8\pi G_N^{4}}{c^4}$ ($G_N^4$ is the Newton's gravitational constant in $4D$ space-time and $c$ is the speed of light) and $g=det\left(g_{\mu\nu}\right)$. Under the assumption of matter distribution throughout the space-time is non homogeneous, the above action is generalised to 
\begin{equation}\label{bal2}
\mathcal{S}_{EH}=\int_{\mathcal{M}}\left(\frac{R}{2K}+L_{\mathscr{M}}\right)\sqrt{-g}~d^4x~~~,
\end{equation}
$L_{\mathscr{M}}\sqrt{-g}$ describes the matter Lagrangian. Actual physical trajectory can be obtained by vanishing the variation of the action \eqref{bal2} with respect to the inverse metric $(g^{\mu\nu})$. Thus we have 
\begin{equation}\label{raj}
\delta \mathcal{S}_{EH}=\int_\mathcal{M}\left(\frac{\delta R}{\delta g^{\mu\nu}}+\frac{R}{\sqrt{-g}}\frac{\delta\sqrt{-g}}{\delta g^{\mu\nu}}+\frac{2K}{\sqrt{-g}}\frac{\delta\left(L_{\mathscr{M}} \sqrt{-g}\right)}{\delta g^{\mu\nu}}\right)\delta g^{\mu\nu} \sqrt{-g}~d^4x=0~~~.
\end{equation}
Since the choice of $\delta g^{\mu\nu}$ is arbitrary, the above equation \eqref{raj} reduces to 
\begin{equation}\label{boka5}
\frac{\delta R}{\delta g^{\mu\nu}}+\frac{R}{\sqrt{-g}}\frac{\delta\sqrt{-g}}{\delta g^{\mu\nu}}=-\frac{2K}{\sqrt{-g}}\frac{\delta\left(L_{\mathscr{M}} \sqrt{-g}\right)}{\delta g^{\mu\nu}}~~~~.
\end{equation}

The variation of scalar curvature $R$ takes the form 
\begin{equation}\label{bal5}
\sqrt{-g}~\delta R=\sqrt{-g}~\delta g^{\mu\nu}R_{\mu\nu}+\sqrt{-g}~g^{\mu\nu}\left\{\nabla_{\rho}\left(\delta\Gamma^{\rho}_{\nu\mu}\right)-\nabla_{\nu}\left(\delta\Gamma^{\rho}_{\mu\rho}\right)\right\}~~~~,
\end{equation}
where the covariant differentiation with respect to the coordinate $x^{\rho}$ is given by
\begin{equation}\nabla_{\rho}A^{\xi}=\frac{\partial A^{\xi}}{\partial x^{\rho}}-\Gamma^{\xi}_{\rho l}A^{l},~\nabla_{\rho}B_{l}=\frac{\partial B_l}{\partial x^\rho}+\Gamma^p_{\rho l}B_p,~0\leq\rho,~\xi,~l,~p\leq 3~~~~.\end{equation}
The variation of Christoffel symbols under the consideration of the relation 
$\partial_{\alpha}\delta g_{\mu\nu}=\nabla_{\alpha}\delta g_{\mu\nu}+\Gamma^{\beta}_{\alpha\mu}\delta g_{\beta\nu}+\Gamma^{\beta}_{\alpha\nu}\delta g_{\beta\mu}$ reduces to 
\begin{equation}
\delta \Gamma^{\mu}_{\alpha\beta}=\frac{1}{2}g^{\mu\nu}\left(\nabla_{\alpha}\delta g_{\beta\nu}+\nabla_{\beta}\delta g_{\alpha\nu}-\nabla_{\nu}\delta g_{\alpha\beta}\right)~~~.
\end{equation}
Second term of RHS of equation \eqref{bal5} reduces to 
$$\sqrt{-g}~g^{\mu\nu}\left\{\nabla_{\rho}\left(\delta\Gamma^{\rho}_{\nu\mu}\right)-\nabla_{\nu}\left(\delta\Gamma^{\rho}_{\mu\rho}\right)\right\}=\nabla_{\rho}\left\{\sqrt{-g}~g^{\mu\nu}~g^{\rho\alpha}\left(\nabla_{\mu}\delta g_{\alpha\nu}-\nabla_{\alpha}\delta g_{\nu\mu}\right)\right\}$$
$$\Rightarrow \sqrt{-g}~g^{\mu\nu}\left\{\nabla_{\rho}\left(\delta\Gamma^{\rho}_{\nu\mu}\right)-\nabla_{\nu}\left(\delta\Gamma^{\rho}_{\mu\rho}\right)\right\}=\partial_{\rho}\left(V^{\rho}\sqrt{-g}\right)~~~~,$$
where, $V^{\rho}=g^{\mu\nu}g^{\rho\alpha}\left(\nabla_{\mu}\delta g_{\alpha\nu}-\nabla_{\alpha} \delta g_{\nu\mu}\right)$. Further, the total derivative under the integration (which is taken over the entire $4D$ manifold $\mathcal{M}$) reduces to zero because at the starting, and at the final point, no variation exists. So from equation \eqref{bal5}, we have 
\begin{equation}\label{bal6}
\frac{\delta R}{\delta g^{\mu\nu}}=R_{\mu\nu}~~~~.
\end{equation}
Again, the variation of determinant of the metric tensor $g_{\mu\nu}$ obeys the relation $ \delta g=g g^{\mu\nu}\delta g_{\mu\nu}$. Thus with the help of the above mentioned relation, we obtain
\begin{equation}\label{bal7}
\frac{1}{\sqrt{-g}}\frac{\delta \sqrt{-g}}{\delta g^{\mu\nu}}=-\frac{1}{2}g_{\mu\nu}~~~.
\end{equation}
On the other hand, RHS of equation \eqref{boka5} can be expanded as
\begin{equation}\label{bal8}
-\frac{2K}{\sqrt{-g}}\frac{\delta\left(L_{\mathscr{M}} \sqrt{-g}\right)}{\delta g^{\mu\nu}}=-\frac{2K}{\sqrt{-g}}\left(\sqrt{-g}~\frac{\delta L_{\mathscr{M}}}{\delta g^{\mu\nu}}+L_{\mathscr{M}}\frac{\delta \sqrt{-g}}{\delta g^{\mu\nu}}\right)=-2K\frac{\delta L_{\mathscr{M}}}{\delta g^{\mu\nu}}+KL_{\mathscr{M}} g_{\mu\nu}=KT_{\mu\nu}~~~.
\end{equation}
Here the stress energy tensor $T_{\mu\nu}$ describes the energy-momentum profile of the core matter of the corresponding stellar models. In this present research work, we have considered the anisotropic distribution of the core matter of the stellar models. Due to such anisotropic distribution of the core matter, the stress energy tensor $T_{\mu\nu}$ can be expressed as \cite{Saha2019}
\begin{equation}
 T_{\mu\nu} = (\rho+ p_t)u_{\mu}u_{\nu} + p_t g_{\mu\nu} + (p_{r}-p_t)v_{\mu}v_{\nu}~~~,~~~u^{\mu} = e^{-\frac{p(r)}{2}}\delta^{\mu}_{0}~,~v^{\mu} = e^{-\frac{q(r)}{2}}\delta^{\mu}_{1}~~~.
\end{equation}
In this regard, energy density, radial component of pressure and tangential component of pressure of the concerned $4-$dimensional stellar models are described by $\rho$, $p_r$ and $p_t$ respectively. Also $p(r)$ and $q(r)$ are metric potentials associated with the interior space-time of the stellar models. On the other hand, the $4$-velocity vector is represented by $u_{\mu}$ and the space-like $4$-vector is represented by $v_{\mu}$. These $4$-vectors satisfy the relations $u_{\mu}u^{\nu} =-1$, $v_{\mu}v^{\nu} =1$, and $u_{\mu}v^{\mu} = 0$. Thus considering equations \eqref{bal6}, \eqref{bal7} and \eqref{bal8}, we have obtained EFE from \eqref{boka5} as
$$G_{\mu\nu}=R_{\mu\nu}-\frac{1}{2}Rg_{\mu\nu}=\frac{8\pi G_N^4}{c^4}T_{\mu\nu}~~~~.\\$$
Considering $G_N^4$ and $c$ in geometric units, i.e.,  $c = G_N^4 = 1$, EFE reduce to 
\begin{equation}\label{bal9}
G_{\mu\nu}=R_{\mu\nu}-\frac{1}{2}Rg_{\mu\nu}=8\pi T_{\mu\nu}~~~~.
\end{equation}

Now the static spherically symmetric space-time which describes the interior stellar structure of the stellar model is of the form
\begin{equation}\label{bal10}
ds_-^2 = -e^{p(r)}dt^{2}+e^{q(r)}dr^{2}+r^2 d\theta^2+r^2 sin^2\theta d\phi^2~~~,
\end{equation}
where, $0\leq\theta\leq\pi ~,~ 0\leq\phi\leq2\pi~$. We have considered the Buchdahl metric-I ansatz for metric potential to obtain a regular singularity free interior solution of the concerned stellar configuration. This type of relativistic fluid sphere solution was first introduced in the research work of Buchdahl \cite{Buchdahl1959}. Subsequently those relativistic fluid sphere solutions which were introduced by Buchdahl have been more generalised in the research works of Vaidya and Tikekar and Durgapal and Bannerji \cite{VT1982,DB1983}. Such Buchdahl class metric ansatz for metric potential can be interpreted as \cite{Sharma2007,Sharma2020,DB1983}
\begin{equation}\label{bal11}
e^{q(r)}=\frac{L\left(1+\tau r^2\right)}{L-\tau r^2}~~~.
\end{equation}
Further, that type of metric potential is extensively used in compact star modelling because it provides researchers a singularity-free interior stellar configuration. Based on such space-time matching conditions, stability and mass–radius characteristics of the concerned stellar model have been depicted significantly in various research works. Sharma and Maharaj have introduced several physically viable compact star models in which the interior stellar configuration of the concerned stellar structures have been speculated within the framework of Buchdahl type relativistic fluid sphere solutions \cite{Sharma2007,Sharma2020}. On the other hand Maurya and his co-researchers have constructed anisotropic stellar model on the basis of Buchdahl metric-I ansatz and also showed the fact that those metric ansatz have incorporated several interior solutions like Vaidya–Tikekar and Finch–Skea solutions \cite{Maurya2019}. Moreover in the research work \cite{Deb2019} physically justified compact star configurations whose interior space-time is depicted by Buchdahl class metric ansatz, have been constructed on the basis of modified gravity theory and Einstein GR and a comparative analysis between the results which are obtained under two gravity theories have been explored significantly. Here $L$ is a dimensionless model constant and it is chosen in such a way that there does not exist any kind of singularity in the interior stellar configuration and $e^{q(r)}>0$. The only possible choice is $L>\tau r^2$ for which \eqref{bal11} is justified. On the other hand, the model constant $\tau$ has dimension of $[length^{-2}]$ and it controls the relation of energy density and compactness of the stellar models with respect to radial coordinate. More precisely $\tau$ plays a very crucial role to obtain realistic and physically viable energy density and compactness of stellar models. The main reason for adopting the Buchdahl metric-I ansatz as an interior metric potential is that it provides a regular interior stellar configuration within the framework of GR \cite{Maurya2019,Deb2019}. Moreover, under the chosen ansatz, the physical parameters (such as energy density, radial and tangential components of pressure etc) required to describe the stellar models exhibit regular and physically reasonable behavior \cite{Sharma2007, Sharma2020}.   

Now under consideration of anisotropic core matter distribution and interior space-time \eqref{bal10}, field equations \eqref{bal9} reduce to 
\begin{equation}
1 - e^{-q} + rq' e^{-q}= 8\pi r^2\rho(r)~~~,
\end{equation}
\begin{equation}
e^{-q}-1 + rp'e^{-q}  = 8\pi r^2 p_{r}(r)~~~\text{and}
\end{equation}
\begin{equation}\label{boka2}
\frac{1}{4}re^{-q}\left\{r(p')^2-2q'+p'(2-rq')+2rp''\right\}=8\pi r^2 p_{t}(r)~~~,
\end{equation}
prime $(')$ denotes the differentiation with respect to the radial coordinate $r$. Now in the next part we will find the metric potential $e^{p(r)}$ for \eqref{bal15}. 
\section{Exterior space-time metric and matching conditions: Neutron Stars in Einstein's Gravity}
In this work, we have considered one parameter EoS, i.e. $p_r=p_r(\rho)$. More precisely, the relation between the energy density ($\rho$) and the radial pressure ($p_r$) of the core matter of the stellar models is not influenced by any other factors. We choose the EoS of the core matter prescribed as \cite{Sharma2007,Lattimer2020} equation \eqref{bal15}. The second and third terms of the prescribed EoS play a very crucial role to describe physically realistic high energy density near the core ($r=0$) of the stellar models \cite{Sharma2007}. 

Now under consideration of such EoS, the interior metric potential $e^{p(r)}$ reduces to 
$$e^{p(r)}=\left\{S+\frac{\tau(L+1)}{384 \pi ^2 L^2}\left(\frac{3 \log \left(r^2 \tau +1\right) \left[128 \pi ^2 A L^2 (L+1)^2+(L+3)^2 \tau  \left\{8 \pi  B L (L+1)+C (L+3) \tau \right\}\right]}{(L+1)^3 \tau }-\frac{4 C \tau(3 L+5)  }{\left(r^2 \tau +1\right)^3}\right.\right.$$
$$~~~~~~~\left.\left.-\frac{3 (L+3) \log \left(L-r^2 \tau \right) \left[64 \pi ^2 A L^2 (L+1)^2+(L+3) \tau  \left\{8 \pi  B L (L+1)+C (L+3) \tau \right\}\right]}{(L+1)^3 \tau }+\frac{384 \pi ^2 L^{3/2} \tanh ^{-1}\left(\frac{r \sqrt{\tau }}{\sqrt{L}}\right)}{\sqrt{\tau }}\right.\right.$$
\begin{equation}\label{bal12}
~~~~\left.\left.-\frac{3 \left\{32 \pi  B L \left(L^2+3 L+2\right)+C\tau (L+3)^3  \right\}}{(L+1)^2 \left(r^2 \tau +1\right)}-\frac{3 \left\{16 \pi  B L (L+1)+C \tau\left(3 L^2+12 L+13\right)  \right\}}{(L+1) \left(r^2 \tau +1\right)^2}-\frac{6 C \tau(L+1)  }{\left(r^2 \tau +1\right)^4}\right)\right\}~~~.
\end{equation}
From equation \eqref{bal12}, it is clear that the metric potential $e^{p(r)}$ behaves regularly in the interior stellar configuration (i.e. in $0\leq r\leq R$; R denotes the stellar radius of the constructed model), and no regularity in $e^{p(r)}$ violates under the consideration of $L>\tau r^2$. Thus, equations \eqref{bal10}, \eqref{bal11}, and \eqref{bal12} together describe the interior space-time of our constructed stellar model with the condition $L>\tau r^2$. In addition, the constant of integration, $S$, can be treated as a model constant, and it plays a very significant role behind the stellar evolution of the corresponding stellar model. 

Now for our constructed stellar model, TOV equations can be speculated as \cite{Deb2021,Saha2019} 
\begin{equation}\label{bal13}
\frac{dM(r)}{dr} = 4\pi r^2\rho(r)~~~\text{and}
\end{equation}
\begin{equation}\label{bal14}
 \frac{dp_{r}}{dr} = -\frac{1}{2}\bigg\{\rho(r)+ p_{r}(r)\bigg\}\frac{dp(r)}{dr}+\frac{2}{r}\bigg\{p_{t}(r)-p_{r}(r)\bigg\}~~~,
\end{equation}
 with the boundary condition $p_r(R)=0$. In this regard,
$\frac{dp_r(r)}{dr}=e^{q(r)}
\left[
8\pi r\,p_r
+\frac{1}{r}\left\{1-e^{-q(r)}\right\}
\right].$
Here, $M(r)$ denotes the total gravitational mass enclosed within a sphere of radius $r$. Using Eqs.~\eqref{bal15}, \eqref{bal10}, \eqref{boka2}, \eqref{bal13}, and \eqref{bal14}, we obtain the energy density $\rho$, the radial pressure $p_r$, and the tangential pressure $p_t$ describing the interior matter distribution of the stellar configuration. Since, in general, $p_r\neq p_t$, the fluid distribution is anisotropic and is characterized by the Bowers--Liang anisotropy factor, $\Delta=p_t-p_r.$
This anisotropy plays a fundamental role in determining the equilibrium, stability, and internal structure of the compact star. Its influence on the physical properties of the stellar configuration is examined in detail in the subsequent physical analysis.

It is evident from the above expressions that these physical quantities depend explicitly on the model parameters $L$, $\tau$, and $S$. Therefore, in the following section, we introduce the exterior Schwarzschild space-time and determine these model parameters by smoothly matching the interior and exterior geometries across the stellar boundary $(r=R)$ through the Darmois--Israel junction conditions.

Here the exterior space-time of the corresponding stellar models is described by Schwarzschild metric which is nothing but the solution of Einstein's field equation. Such space-time in $4-$dimension is defined by \cite{Saha2019}
\begin{equation}
ds_+^{2} = -\left(1-\frac{2M(r=R)}{r}\right)dt^2+\left(1-\frac{2M(r=R)}{r}\right)^{-1}dr^2+r^2 d\theta^2+r^2 sin^2\theta d\phi^2~~~,
\end{equation}
where, $0\leq\theta\leq\pi ~,~ 0\leq\phi\leq2\pi~$. Now the matching condition of the interior and exterior space-time metric of the concerned stellar models at the boundary $(r=R)$ can be interpreted as \cite{Maurya2017,Deb2021,Saha2019}
\begin{equation}
g_{tt}^{-}= g_{tt}^{+},~g_{rr}^{-}= g_{rr}^{+}~\text{and}~p_r(R)=0~~~.
\end{equation}
Here ${\bf -}$ and ${\bf +}$ represent interior and exterior metric potentials respectively. Now for our constructed stellar model, matching condition at the boundary reduces to 
$$e^{p(R)}= 1-\frac{2M(r=R)}{R},~e^{q(R)}= \bigg\{1-\frac{2M(r=R)}{R}\bigg\}^{-1}~\text{and}~~p_r(R)=0~~~.$$
From the foregoing analysis, it is evident that the metric potentials, $e^{p(R)}$ and $e^{q(R)}$, together with the radial pressure at the stellar boundary, $p_r(R)$, depend explicitly on the model parameters $L$, $\tau$, and $S$. The corresponding boundary conditions provide three linearly independent equations for these three unknown parameters. Consequently, the matching conditions uniquely determine the values of $L$, $\tau$, and $S$ for each compact-star candidate considered in the present work.

The resulting parameter values correspond to physically acceptable stellar configurations that remain regular throughout the interior region, i.e., for $0\leq r\leq R$. Once these parameters are determined, the total gravitational mass enclosed within the stellar radius $R$ is obtained from Eq.~\eqref{bal13} by evaluating $M=M(R)$ as

\begin{equation}
M(r=R)=\int_{r=0}^{r=R}4\pi r^2\rho(r)dr~~~.
\end{equation}
From the mathematical formulation of the proposed stellar model, it is evident that the dimensionless model parameter $L$ plays a crucial role in determining the regularity and physical acceptability of the interior space-time. In particular, the structural properties of the compact star, including its mass--radius relation, are strongly influenced by the choice of this parameter. Consequently, we investigate the dependence of the mass--radius characteristics of the proposed stellar model on the model parameter $L$ for several observed neutron stars.

Furthermore, the radial distributions of the relevant physical quantities, including the energy density, radial and tangential pressures, anisotropy, and enclosed mass, are analyzed as functions of the radial coordinate $r$. In addition, various physical properties, equilibrium conditions, and hydrodynamic stability criteria are examined in detail to assess the physical viability and astrophysical applicability of the proposed stellar model for the compact-star candidates considered in this work.

\section{Physical analysis: Results Obtained from the Model}
In this section, we investigate the structural properties, hydrodynamic stability, equilibrium configuration, and various physical characteristics of the proposed stellar model. For the graphical and numerical analyses, we consider four well-studied neutron stars: $PSR~J0614-3329$ ($M_{Obs}=1.44\pm0.07,M_{\odot}$, $R_{Rep}=10.29\pm1.01,\mathrm{km}$), $PSR~J0437-4715$ ($M_{Obs}=1.42\pm0.04,M_{\odot}$, $R_{Rep}=11.36\pm0.95,\mathrm{km}$), $Vela~X$-$1$ ($M_{Obs}=1.77\pm0.08,M_{\odot}$, $R_{Rep}=11.75\pm0.85,\mathrm{km}$), and $PSR~J1614$-$2230$ ($M_{Obs}=1.97\pm0.04,M_{\odot}$, $R_{Rep}=13.00\pm2.00,\mathrm{km}$) \cite{Sharma2007,Mauviard2025,Choudhury2024,Abubekerov2008,Elebert2009}. Here, $M_{Obs}$ and $R_{Rep}$ denote the observationally inferred mass and the reported stellar radius, respectively.

For each compact-star candidate, a representative mass is selected within its observationally constrained mass interval. Using the proposed stellar model, we then determine the corresponding stellar radius and verify that the predicted value lies within the reported observational range. Once a physically acceptable mass--radius pair is obtained, the associated model parameters are determined by applying the boundary matching conditions. The resulting numerical values of these parameters are listed in Table~\ref{tab:maxmassk}.
   
\begin{table}
\small
\centering
\caption{\textbf{model constants of the above mentioned neutron stars for constructed stellar model}}
\label{tab:maxmassk}
\begin{center}
\vspace{0.1 in}
\begin{tabular}{
|p{0.4cm}|p{3.5cm}|p{1.8cm}|p{2.0cm}|p{1.2cm}|p{1.2cm}|p{1.2cm}|p{1.2cm}|p{1.2cm}|}
\hline
\multicolumn{9}{|c|}
{model constants for the neutron stars}\\
\hline
Sl. No.&Neutron Star&$M_{Obs}(M_{\odot})$&$R_{Rep}[km]$& $M[M_{\odot}]$& $R[km]$& $L[km^{0}]$&$\tau [km^{-2}]$& $S [km^{0}]$\\
\hline
$01.$ & $PSR~J1614-2230$ \cite{Elebert2009}& $1.97\pm 0.04$  &  $13.00\pm 2.00$  &  $1.97$ & $13.06$  & $9.7123$&$0.0497$& $0.7502$\\
\hline
$02.$ & $Vela~X-1$\cite{Abubekerov2008} & $1.77\pm 0.08$&$11.75\pm 0.85$&$1.77$ & $12.56$& $8.5261$&$0.0481$& $0.7204$\\
\hline
$03.$&$PSR~J0614-3329$\cite{Mauviard2025}& $1.44\pm 0.07$  &  $10.29\pm 1.01$  &  $1.44$ & $11.01$& $6.7662$&$0.0311$& $0.6493$\\
\hline
$04.$&$PSR~J0437-4715$\cite{Choudhury2024}& $1.42\pm 0.04$  &  $11.36\pm 0.95$  &  $1.41$ & $10.73$& $6.7618$&$0.0320$& $0.6572$\\
\hline
\end{tabular}
\end{center}
\vspace{2mm}
\footnotesize
\begin{justify}
\textit{\bf{Table~1:}} presents the model parameters $L$, $\tau$, and $S$ corresponding to the compact stars (A) $PSR~J1614-2230$, (B) $Vela~X-1$, (C) $PSR~J0614-3329$, and (D) $PSR~J0437-4715$. The obtained numerical values of the model parameters demonstrate that all stellar configurations satisfy the regularity condition $L>\tau r^{2}$ throughout the stellar interior. In addition, the third and fourth columns list the observationally inferred mass range, $M_{\rm Obs}/M_{\odot}$, and the corresponding reported radius, $R_{\rm Rep}$ (km), respectively. From each observationally inferred mass interval, a representative mass, $M/M_{\odot}$, is selected (fifth column), while the corresponding stellar radius, $R$ (sixth column), is determined from the theoretical mass--radius relation obtained within the framework of the present model. The representative mass is chosen such that the resulting stellar configuration simultaneously satisfies the regularity condition $L>\tau r^{2}$ and the boundary condition of the TOV equation, namely, $p_r(R)=0$.
\end{justify}
\end{table}
\begin{itemize}

\item Figure~1 illustrates the radial variation of the energy density $\rho(r)$ inside the neutron-star configurations considered in the present work. It is evident that the energy density decreases monotonically with increasing radial coordinate, i.e.,
\[
\frac{d\rho}{dr}<0,\qquad 0\le r\le R,
\]
which is one of the fundamental physical requirements for realistic compact-star models.

~~~At the stellar center ($r=0$), the energy density attains its maximum value owing to the extreme gravitational compression of matter in the core. Under such conditions, matter exists in an ultra-dense state supported primarily by neutron degeneracy pressure together with strong nuclear interactions. As the radial distance increases, the gravitational field becomes progressively weaker, leading to a continuous decrease in the matter density toward the stellar surface ($r=R$), where the energy density reaches its minimum value.

~~~Such a monotonic density profile is consistent with hydrostatic equilibrium, in which the inward gravitational attraction is balanced by the outward pressure gradient governed by the TOV equation. Furthermore, the absence of any density discontinuities or unphysical oscillations confirms the regularity and physical acceptability of the interior solution.

~~~For the stellar configurations considered in the present work, the more massive neutron stars exhibit comparatively higher energy densities throughout their interiors than the less massive configurations. This behavior reflects the stronger gravitational field of massive compact stars, which requires a higher central matter concentration to maintain hydrostatic equilibrium. Consequently, the energy-density profiles corresponding to the higher-mass stars lie above those of the lower-mass configurations over the entire interior region.

~~~Overall, the behavior displayed in Fig.~1 is consistent with the expected properties of realistic neutron-star interiors and further supports the physical viability of the proposed stellar model.

\end{itemize}\begin{itemize}

\item Figure~2 illustrates the radial variation of the radial pressure $p_r(r)$ within the interior of the neutron-star configurations considered in the present work. It is evident that the radial pressure decreases monotonically with increasing radial coordinate, i.e.,
\[
\frac{dp_r}{dr}<0,\qquad 0\leq r<R,
\]
which is one of the essential physical requirements for realistic compact-star models. The pressure profiles remain smooth and positive throughout the stellar interior, indicating the absence of any discontinuity or unphysical behavior.

~~~At the stellar center ($r=0$), the radial pressure attains its maximum value owing to the extreme gravitational compression of matter in the core region. As the radial distance increases, the inward gravitational attraction becomes progressively weaker, resulting in a continuous decrease in the pressure toward the stellar surface. Finally, the radial pressure vanishes smoothly at the boundary, i.e., $p_r(R)=0$, satisfying the physical boundary condition required for matching the interior solution to the exterior Schwarzschild space-time through the Darmois--Israel junction conditions.

~~~Such a monotonic pressure profile is consistent with the condition of hydrostatic equilibrium, where the inward gravitational force is balanced by the outward pressure gradient governed by the TOV equation. The smooth behavior of the pressure distribution further confirms the regularity and physical acceptability of the interior solution.

~~~A comparative analysis of the pressure profiles reveals that, for the stellar configurations considered in this work, the more massive neutron stars sustain comparatively higher radial pressures throughout their interiors than the lower-mass configurations. This behavior reflects the stronger gravitational field associated with massive compact stars, which requires a larger internal pressure to counterbalance gravitational collapse and maintain equilibrium. Consequently, the radial-pressure profiles corresponding to the more massive stars consistently lie above those of the less massive configurations over the entire interior region.

~~~Overall, the behavior displayed in Figure~2 agrees well with the expected physical characteristics of realistic neutron-star interiors and further supports the physical viability of the proposed stellar model.

\end{itemize}
\begin{itemize}

\item Figure~3 illustrates the radial variation of the tangential pressure $p_t(r)$ within the interior of the neutron-star configurations considered in the present work. It is evident that the tangential pressure remains positive throughout the stellar interior, i.e.,
\[
p_t(r)>0,\qquad 0\le r\le R,
\]
indicating a physically acceptable matter distribution without any unphysical behavior.

~~~The tangential pressure decreases smoothly and monotonically with increasing radial coordinate, satisfying
\[
\frac{dp_t}{dr}<0,\qquad 0\le r\le R.
\]
It attains its maximum value at the stellar center ($r=0$), where the matter is subjected to extreme gravitational compression, and gradually decreases toward the stellar surface as the gravitational field weakens. Such a monotonic profile is consistent with the expected behavior of realistic compact-star interiors.

~~~A comparative analysis of the profiles reveals that, for the stellar configurations considered in the present work, the more massive neutron stars sustain comparatively higher tangential pressures throughout their interiors than the lower-mass configurations. This behavior reflects the stronger gravitational field associated with massive compact stars, which requires greater internal stresses to maintain hydrostatic equilibrium.

~~~A direct comparison between Figures~2 and~3 further shows that
\[
p_t>p_r,\qquad 0<r\le R,
\]
throughout the stellar interior. Consequently, the anisotropy factor,
\[
\Delta=p_t-p_r,
\]
remains positive everywhere inside the star. Such positive anisotropy generates an outward-directed anisotropic force that acts against gravity and thereby contributes to supporting more massive compact-star configurations.

~~~The smooth and regular behavior of the tangential pressure, together with the consistently positive anisotropy throughout the stellar interior, demonstrates that anisotropic stresses are naturally incorporated into the present stellar model. This feature plays an important role in determining the equilibrium and structural properties of the compact-star configurations considered in this work.

~~~Overall, the behavior displayed in Figure~3 is consistent with the expected physical characteristics of anisotropic neutron-star interiors and further supports the physical viability of the proposed stellar model.

\end{itemize}
\begin{itemize}

\item Figure~4 illustrates the radial variation of the anisotropy factor,
\[
\Delta(r)=p_t(r)-p_r(r),
\]
for the neutron-star configurations considered in the present work. It is evident that
\[
\Delta(0)=0,
\]
indicating that the fluid distribution is isotropic at the stellar center. This is an essential requirement for obtaining a regular interior solution. As the radial coordinate increases, the anisotropy factor grows smoothly and monotonically, attaining its maximum value near the stellar surface.

~~~A careful examination of the profiles shows that the increase in $\Delta$ is gradual in the central region but becomes more pronounced toward the outer layers of the star. This behavior indicates that anisotropic stresses remain relatively weak near the core and become increasingly significant as one approaches the stellar surface.

~~~Since the anisotropic force is given by
\[
F_a=\frac{2\Delta}{r},
\]
its magnitude remains very small in the vicinity of the stellar center and increases progressively toward the boundary. Because $\Delta>0$ throughout the stellar interior, the corresponding anisotropic force is directed outward and acts in opposition to gravity. Consequently, it assists the pressure gradient in maintaining hydrostatic equilibrium and contributes to the supporting compact stellar configurations.

~~~A comparison of the anisotropy profiles further reveals that the more massive neutron-star configurations exhibit comparatively larger values of $\Delta$ throughout their interiors than the lower-mass configurations. This behavior reflects the increasing importance of anisotropic stresses in supporting compact stars with stronger gravitational fields.

~~~The smooth, positive, and monotonically increasing behavior of the anisotropy factor demonstrates that the present stellar model remains free from any unphysical irregularities throughout the interior. Overall, the behavior displayed in Figure~4 is consistent with the expected properties of physically acceptable anisotropic compact-star models and further supports the physical viability of the proposed stellar configuration.

\end{itemize}

\begin{figure}[H]
\begin{center}
\includegraphics[keepaspectratio=true,height=3.5in, width=4.5in]{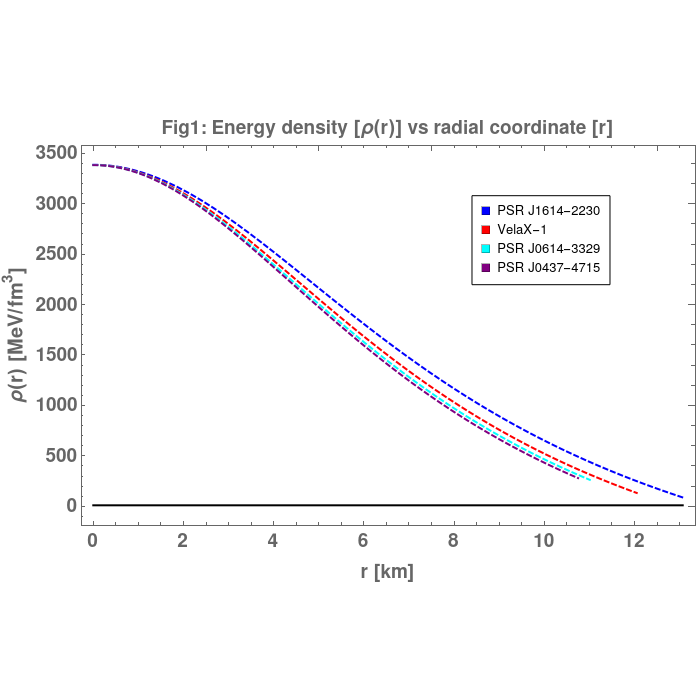}~~~~\includegraphics[keepaspectratio=true,height=3.5in, width=4.5in]{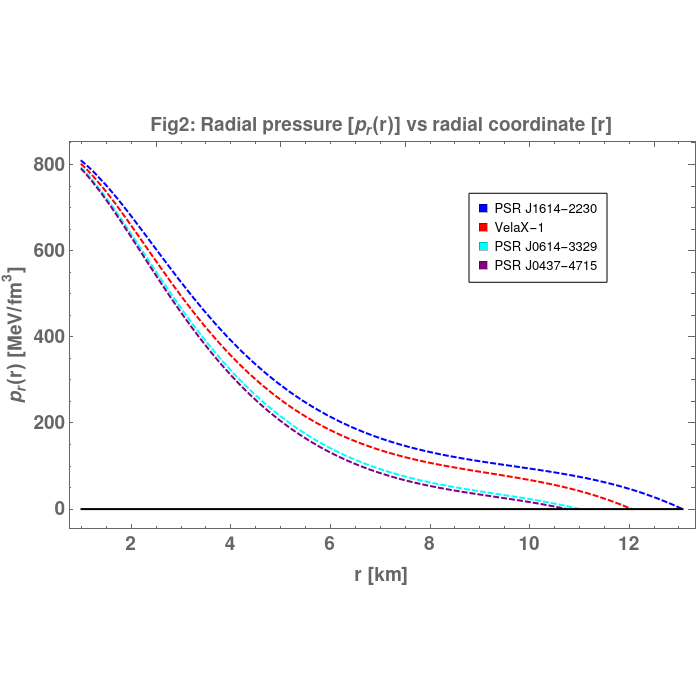}\\
\includegraphics[keepaspectratio=true,height=3.5in, width=4.5in]{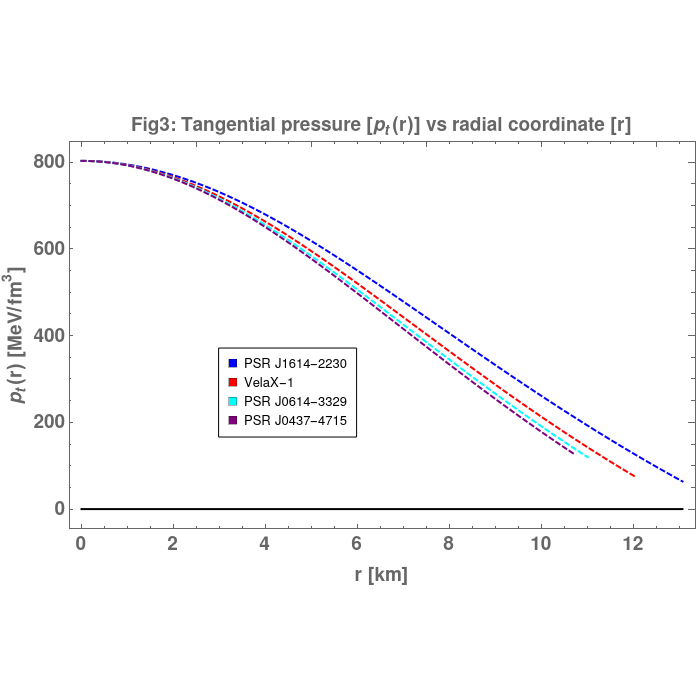}~~~~\includegraphics[keepaspectratio=true,height=3.5in, width=4.5in]{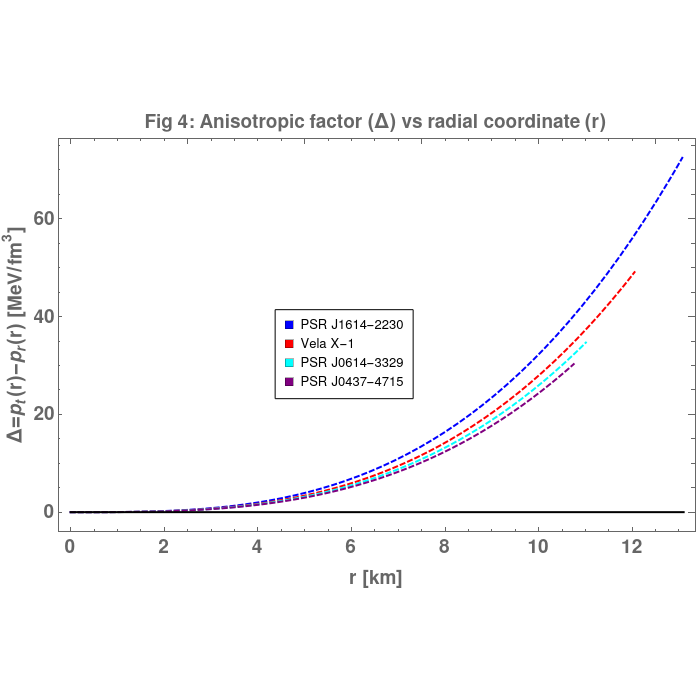}\\
\begin{justify}
Figures: 1--4 represent the profiles of the physical parameters, namely the energy density ($\rho$), radial pressure ($p_r$), tangential pressure ($p_t$), and the anisotropic factor ($\Delta$), as functions of the radial coordinate ($r$) for the neutron stars (A) $PSR~J1614-2230$, (B) $Vela~X-1$, (C) $PSR~J0614-3329$, and (D) $PSR~J0437-4715$ under the constructed models. The stellar surface for each configuration is determined by the physical boundary condition $p_r(R)=0$ (boundary condition of the TOV equation; from Fig:~2). Accordingly, the radial pressure vanishes at $R=13.0729~\mathrm{km}$ for $PSR~J1614-2230$, $R=12.0518~\mathrm{km}$ for $Vela~X-1$, $R=11.0148~\mathrm{km}$ for $PSR~J0614-3329$, and $R=10.7581~\mathrm{km}$ for $PSR~J0437-4715$. Since the interior solution is physically meaningful only within the stellar boundary ($0 \leq r \leq R$), the profiles of the energy density ($\rho$), tangential pressure ($p_t$), and the anisotropic factor ($\Delta$) are presented only up to their respective stellar surfaces. This avoids extending the physical variables into the unphysical exterior region beyond the compact star.\\
\end{justify}
\end{center}
\end{figure}

\begin{itemize}

\item Figures~5(a--d) illustrate the three-dimensional variation of the enclosed gravitational mass $M(r)$ as a function of the radial coordinate $r$ and the model parameter $L$ for the neutron-star configurations considered in the present work. The primary purpose of presenting the three-dimensional profiles is to provide a clear visualization of the simultaneous dependence of the enclosed mass on both the radial coordinate and the model parameter. In particular, the horizontal level surface corresponding to the maximum estimated mass serves as a convenient visual reference for identifying the limiting mass attained by each stellar configuration. The profiles clearly demonstrate that $M(r)$ increases monotonically with $r$, starting from $M(0)=0$ at the stellar center and attaining its maximum value at the stellar surface ($r=R$). This behavior is consistent with the standard requirement of a physically acceptable mass function, ensuring a regular and well-defined matter distribution throughout the stellar interior.

~~~A notable feature emerging from these plots is the smooth and continuous growth of $M(r)$ without any irregularity throughout the interior region, indicating a regular accumulation of matter from the stellar center toward the surface. The three-dimensional representation further illustrates that this monotonic increase is preserved for all admissible values of the model parameter $L$, while the level surface provides a clear indication of the maximum mass associated with each neutron-star configuration. Thus, the three-dimensional visualization complements the quantitative results presented in Table~\ref{tab:maxmass} and offers a more intuitive understanding of the dependence of the enclosed mass on both $r$ and $L$.

~~~From the graphical analysis, it is evident that the maximum estimated mass ($M_{\mathrm{Est}}$) obtained from the constructed model exceeds the corresponding observed mass ($M_{\mathrm{Obs}}$) for all the considered neutron stars, as summarized in Table~\ref{tab:maxmass}. This systematic enhancement in mass can be attributed to the combined effects of hydrostatic pressure and anisotropic stresses within the stellar interior. In particular, the presence of a positive anisotropy factor ($p_t>p_r$) introduces an additional outward-directed force that effectively counteracts gravitational collapse and enables the star to support more massive equilibrium configurations.

~~~This result is physically significant, as it indicates that anisotropy can play an important role in extending the maximum mass limit of neutron stars beyond that predicted by purely isotropic models. Such an effect has been discussed in earlier studies of anisotropic compact stars \cite{Bowers1974,Herrera1997}, where anisotropic stresses were shown to enhance the mass-bearing capacity of relativistic stellar configurations.

~~~Moreover, the obtained mass range ($\sim2.2$--$2.8\,M_{\odot}$) is particularly noteworthy, as it lies close to or slightly above the observationally inferred upper limits of neutron-star masses. For example, massive pulsars such as PSR~J1614$-$2230 and PSR~J0740$+$6620 possess masses close to $2\,M_{\odot}$ \cite{Cromartie2020,Demorest2010}, while gravitational-wave observations of GW170817 suggest an upper limit of approximately $2.2$--$2.3\,M_{\odot}$ depending on the adopted EoS \cite{Abbott2017,Abbott2018}.

~~~Within the framework of the present model, the slightly larger maximum masses can be interpreted as a consequence of the additional support provided by positive anisotropic stresses. This suggests that anisotropy may provide an effective mechanism for supporting massive neutron stars without requiring an extremely stiff equation of state.

~~~Another important outcome is the preservation of the observed mass hierarchy among the selected neutron-star configurations, with comparatively heavier observed stars yielding larger estimated maximum masses within the present model.

~~~Therefore, the principal outcomes of the present analysis are:
\\
(i) the demonstration of a smooth and physically acceptable mass function satisfying the fundamental requirements of relativistic stellar structure,
\\
(ii) the visualization of the monotonic increase of the enclosed mass with the radial coordinate and the model parameter through the three-dimensional representation,
\\
(iii) the prediction of enhanced maximum mass limits arising from anisotropic stresses,
\\
(iv) the consistency of the estimated mass range with current astrophysical observations, and
\\
(v) the indication that pressure anisotropy can effectively support massive neutron-star configurations without invoking extremely stiff equations of state.

~~~Overall, the mass--radius characteristics presented in Figures~5(a--d), together with the numerical results summarized in Table~\ref{tab:maxmass}, demonstrate that the proposed stellar model provides a physically consistent and observationally relevant description of compact neutron stars.
\end{itemize}\begin{table}[ht]
\centering
\caption{\textbf{mass-radius characteristics of neutron stars for our constructed stellar model}}
\label{tab:maxmass}
\begin{center}
\vspace{0.1 in}
\begin{tabular}{
|p{0.4cm}|p{3.8cm}|p{1.8cm}|p{2.0cm}|p{1.2cm}|p{1.2cm}|p{1.7cm}|p{1.7cm}|}
\hline
\multicolumn{8}{|c|}
{mass-radius relation of neutron stars}\\
\hline
Sl. No.&Neutron Star&$M_{Obs}(M_{\odot})$&$R_{Rep}[km]$& $M[M_{\odot}]$& $R[km]$& $M_{Est}(M_{\odot})$&$R_{Est}[km]$\\
\hline
$01.$ & $PSR~J1614-2230$ \cite{Elebert2009}& $1.97\pm 0.04$  &  $13.00\pm 2.00$  &  $1.97$ & $13.06$  & $2.7645$& $13.0729$\\
\hline
$02.$ & $Vela~X-1$\cite{Abubekerov2008} & $1.77\pm 0.08$&$11.75\pm 0.85$&$1.77$ & $12.56$& $2.6074$& $12.0518$ \\
\hline
$03.$&$PSR~J0614-3329$\cite{Mauviard2025}& $1.44\pm 0.07$  &  $10.29\pm 1.01$  &  $1.44$ & $11.01$& $2.3806$& $11.0148$\\
\hline
$04.$&$PSR~J0437-4715$\cite{Choudhury2024}& $1.42\pm 0.04$  &  $11.36\pm 0.95$  &  $1.41$ & $10.73$&$2.2411$& $10.7581$\\
\hline
\end{tabular}
\end{center}
\vspace{2mm}
\footnotesize
\begin{justify}
\textit{\bf{Table~2:}} Presents the maximum estimated mass ($M_{\rm Est}/M_{\odot}$) of the compact stars (A) $PSR~J1614-2230$, (B) $Vela~X-1$, (C) $PSR~J0614-3329$, and (D) $PSR~J0437-4715$, together with the corresponding stellar radius, $R_{\rm Est}$ (km), at which each configuration attains its maximum stable mass within the framework of the proposed stellar model. The numerical results indicate that the predicted maximum masses of all four compact stars fall within the lower mass-gap region. This suggests that the present model is capable of supporting massive neutron star configurations with masses extending into the lower mass-gap regime. 

\end{justify}
\end{table}
\newpage
\begin{figure}
\begin{center}
\includegraphics[keepaspectratio=true,height=3.6in, width=7.5in]{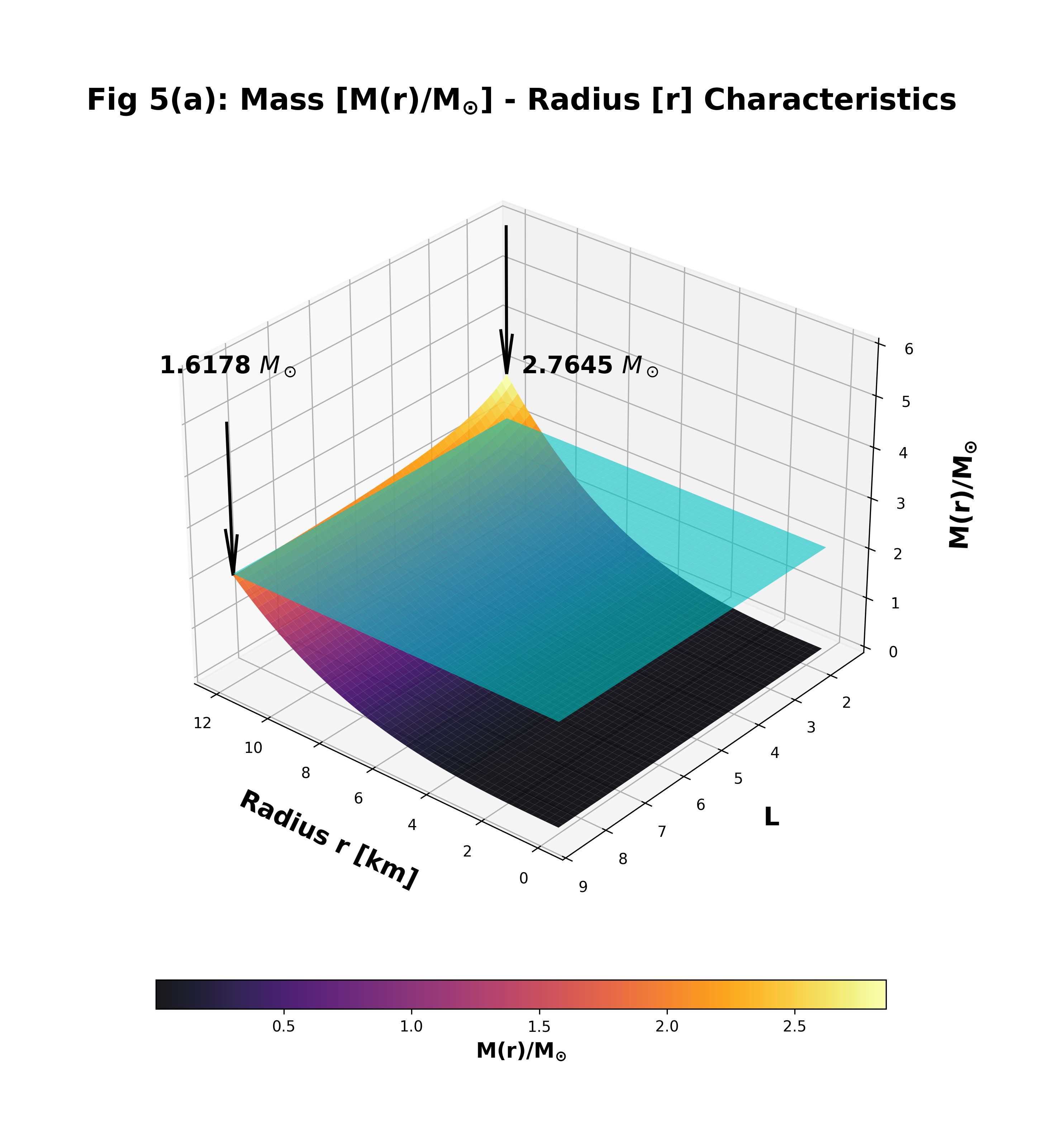}~~~~\includegraphics[keepaspectratio=true,height=3.6in, width=7.5in]{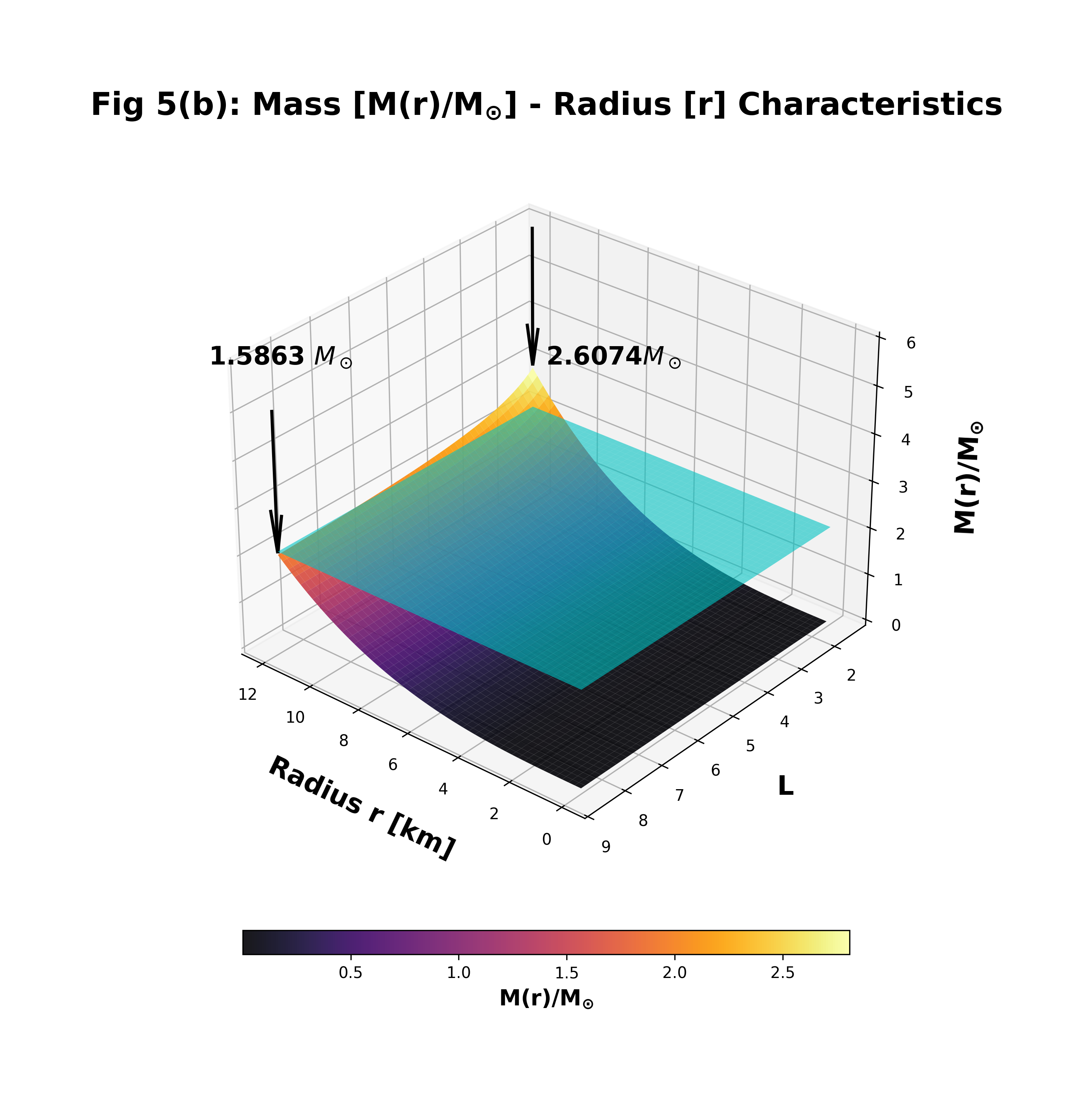}\\
\includegraphics[keepaspectratio=true,height=3.6in, width=7.5in]{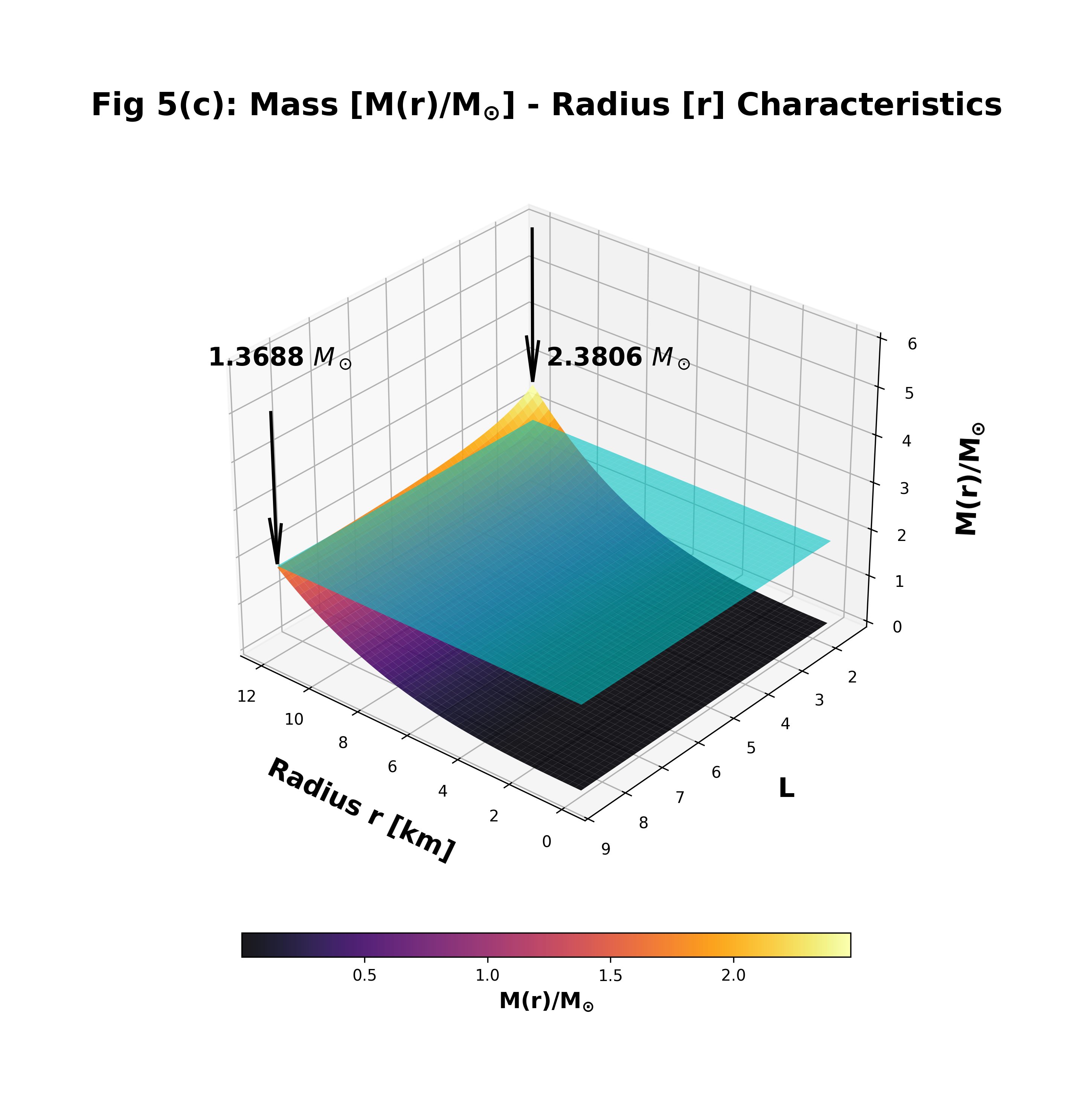}~~~~\includegraphics[keepaspectratio=true,height=3.6in, width=7.5in]{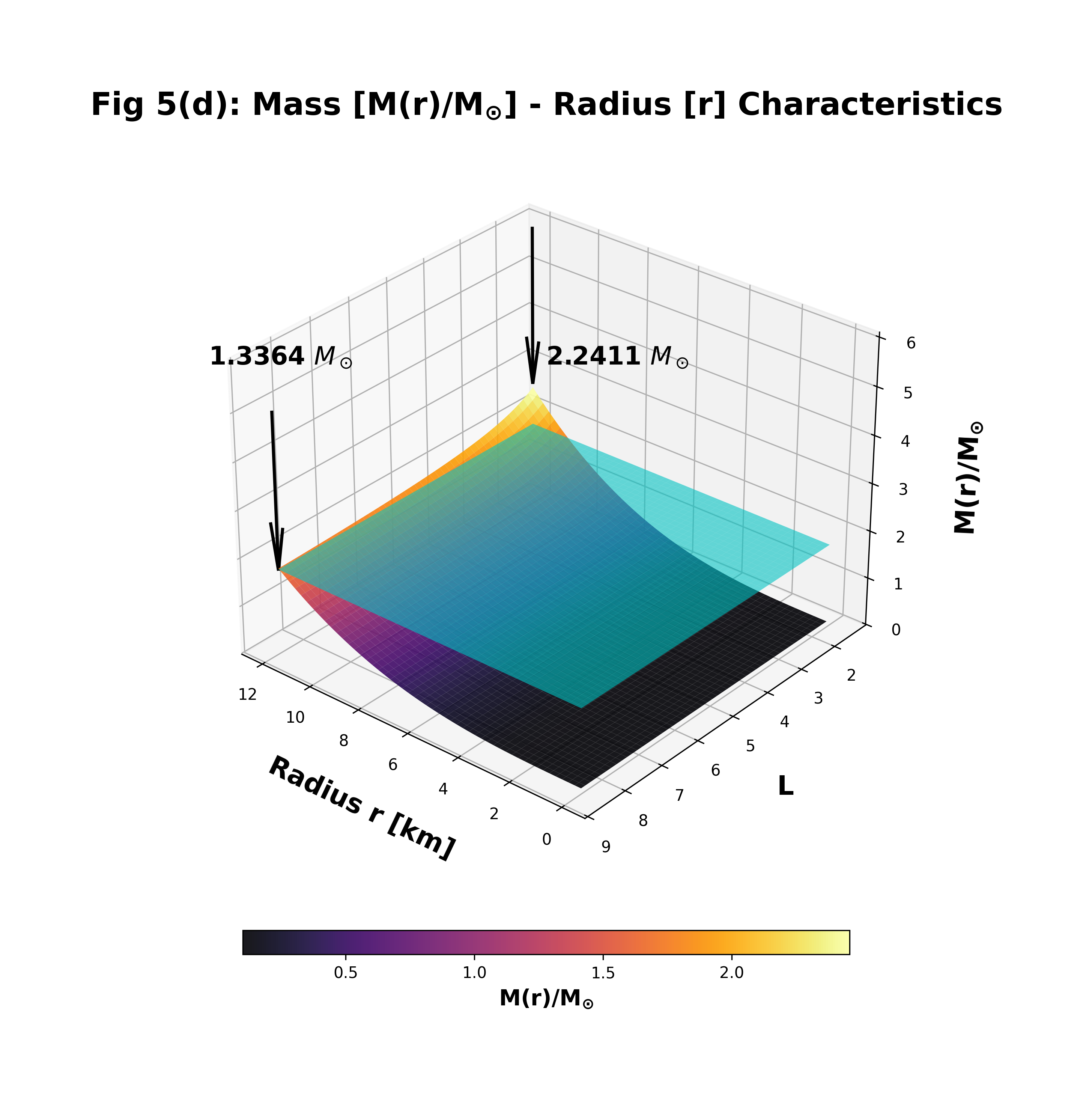}\\
\begin{justify}
Figures: 5(a-d) depict the three-dimensional variation of the enclosed mass, $M(r)/M_{\odot}$, as a function of the radial coordinate, $r$ (km), and the model parameter, $L$ ($\mathrm{km}^{-2}$), for the neutron stars (A) $PSR~J1614-2230$, (B) $Vela~X-1$, (C) $PSR~J0614-3329$, and (D) $PSR~J0437-4715$ within the framework of the proposed model. To facilitate a clearer visualization of the mass evolution with respect to the radial coordinate and the model parameter, each three-dimensional mass surface is intersected by a horizontal plane corresponding to $M=2.06\,M_{\odot}$ for $PSR~J1614-2230$, $M=2.03\,M_{\odot}$ for $Vela~X-1$, $M=1.79\,M_{\odot}$ for $PSR~J0614-3329$, and $M=1.75\,M_{\odot}$ for $PSR~J0437-4715$. The resulting intersection curves provide a clearer representation of the monotonic increase of the enclosed mass with the radial coordinate and simultaneously highlight the maximum estimated mass attained by each compact star within the considered parameter space.\\
\end{justify}
\end{center}
\end{figure}
\begin{itemize}
\item When a neutron star forms a binary system with another compact object, the external tidal gravitational field perturbs its equilibrium configuration, causing a deviation from perfect spherical symmetry. Such tidal deformations are conveniently described within the framework of linear perturbation theory by introducing a small perturbation $\eta_{\mu\nu}$ to the background interior metric tensor $g^{-}_{\mu\nu}$. Consequently, the perturbed interior metric takes the form  \cite{Hinderer2008}
$$(g^-_{\mu\nu})_{modified}=g^-_{\mu\nu}+\eta_{\mu\nu}~~~,$$
represents the perturbed space-time. Here the small perturbation $\eta_{\mu\nu}$ is of the form  \cite{Binnington2009}
\[
\eta_{\mu\nu} =
\begin{pmatrix}
e^{p(r)} \mathcal{H}(r) & 0 & 0 & 0 \\
0 & e^{q(r)} \mathcal{H}(r) & 0 & 0 \\
0 & 0 & r^2 \mathcal{K}(r) & 0 \\
0 & 0 & 0 & r^2 \sin^2\theta \, \mathcal{K}(r)
\end{pmatrix}
P_2(\cos\theta)~~~~.
\]
In this regard $\mathcal{H}(r)$ and $\mathcal{K}(r)$ are perturbation function associated with the radial coordinate $r$ and $P_2(\cos\theta)$ is the Legendre polynomial for the particular choice $l=2$. Under this consideration the static spherically symmetric interior space-time \eqref{bal10} reduces to  \cite{Hinderer2008,Binnington2009}   
\[
ds_{-}^2 =
- e^{p(r)}\left[1 + \mathcal{H}(r) P_2(\cos\theta)\right] dt^2
+ e^{q(r)}\left[1 + \mathcal{H}(r) P_2(\cos\theta)\right] dr^2
+ r^2 \left[1 + \mathcal{K}(r) P_2(\cos\theta)\right]
\left(d\theta^2 + \sin^2\theta d\phi^2\right)~~~.
\]
Moreover the radial perturbation function $\mathcal{K}(r)$ can be expressed as function of $\mathcal{H}(r)$ i.e. $\mathcal{K}(r)=f\left(\mathcal{H}(r),\mathcal{H}'(r)\right)$. Now on the basis of this construction and general relativistic equations we have obtained a differential equation which is of the form \cite{Hinderer2008}
\begin{equation}\label{chal1}
r\frac{dy(r)}{dr}+y^2+y(r)\mathcal{F}(r)+r^2\mathcal{Q}(r)=0~~~,
\end{equation}
where $y(r)=\frac{r\mathcal{H}'(r)}{\mathcal{H}(r)}$. Moreover $\mathcal{F}(r)$ and $\mathcal{Q}(r)$ are depends on energy density ($\rho$), radial pressure component ($p_r$) and mass parameter ($M(r)$) of the stellar models and can be expressed as \cite{Hinderer2008,Binnington2009} 
\[
\mathcal{F}(r)=\frac{r-4\pi r^3\left\{\rho(r)-p_r(r)\right\}}{r-2M(r)}~~~\text{and}\]
\[
\mathcal{Q}(r)=\frac{4\pi r\left\{5\rho(r)+9p_r(r)+
\frac{\rho(r)+p_r(r)}{v_r^2}\right\}}{\left\{r-2M(r)\right\}}
-\frac{6}{r^2}
-\frac{4\left\{M(r)+4\pi r^3 p_r(r)\right\}^2}{r^2\left\{r-2M(r)\right\}^2}~~~.
\]
Here the radial sound velocity $v_r^{2}=\frac{dp_r}{d\rho}~$. Now integrating the above differential equation \eqref{chal1}, from $r=0$ to $r=R$, we have obtained $y_F=y(R)$. Right now, we have introduced the tidal love number $\kappa$ as a function of compactness ($u=\frac{M(r)}{r}$) and $y_{F}$. Such love number can be represented as
$$\kappa =\frac{8u^5}{5}\Bigg(1-2u\Bigg)^2\Bigg\{2u(y_F-1)-y_F+2\Bigg\}\Bigg[2u\Bigg\{6-3y_F+3u(5y_F-8)\Bigg\}
+4u^3\Bigg\{13-11y_F+u(3y_F-2)$$
\begin{equation}\label{arpann}
~~~~~~~~~~~~~~~~~~~~~~~~~~~~~~~~~~~~~~~~~~~~~~~~~~~~~~+2u^2(1+y_F)\Bigg\}+3(1-2u)^2\Bigg\{2-y_F+2u(y_F-1)\Bigg\}\ln(1-2u)\Bigg]^{-1}.
\end{equation}
Now with help of $\kappa$, tidal deformability ($\Lambda$) of the corresponding stellar model can be interpreted as
$$\Lambda=\frac{2}{3}\kappa u^{-5}~~~.$$
\begin{figure}[H]
\begin{center}
\includegraphics[keepaspectratio=true,height=2.4in, width=3.5in]{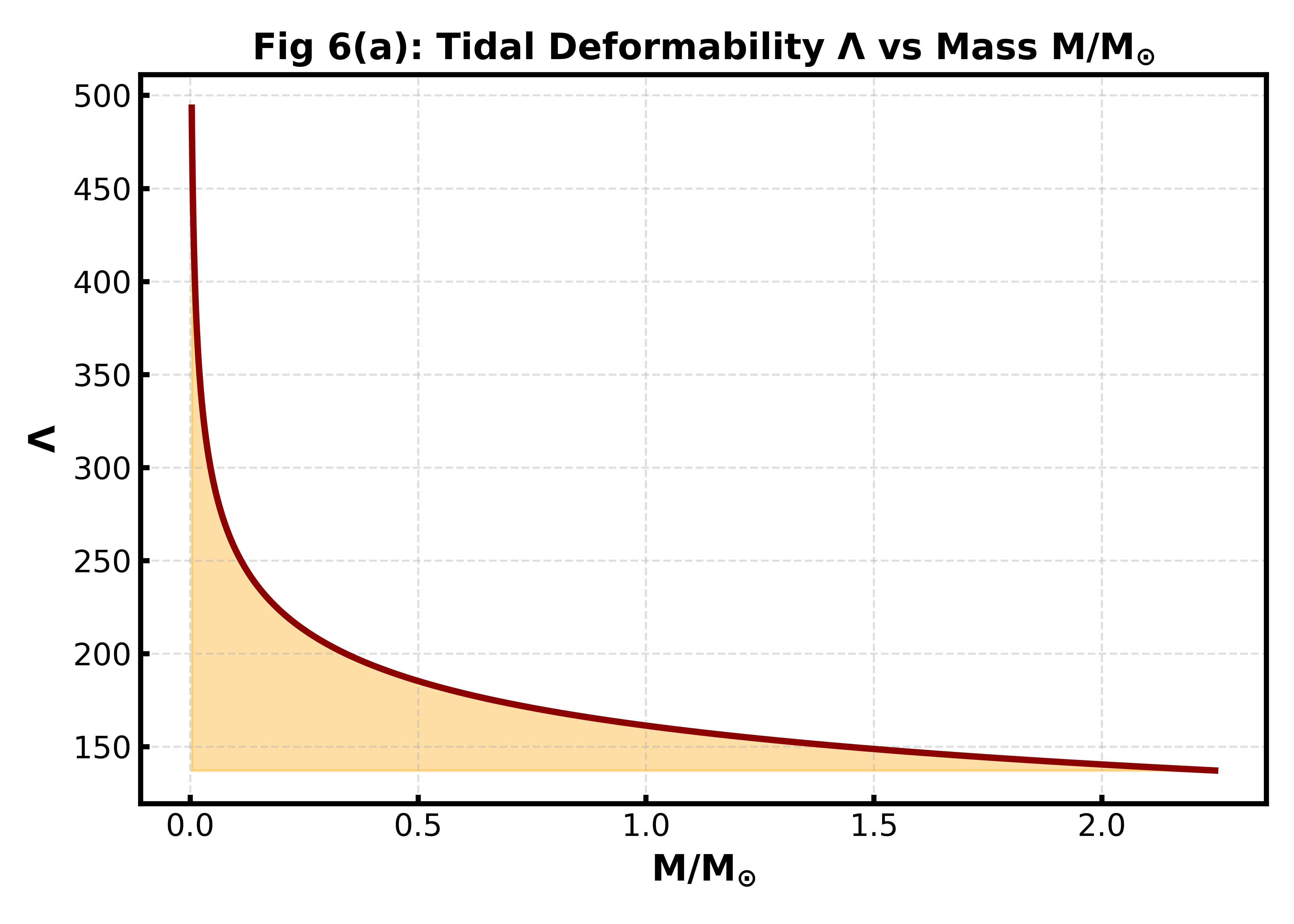}~~~~\includegraphics[keepaspectratio=true,height=2.4in, width=3.5in]{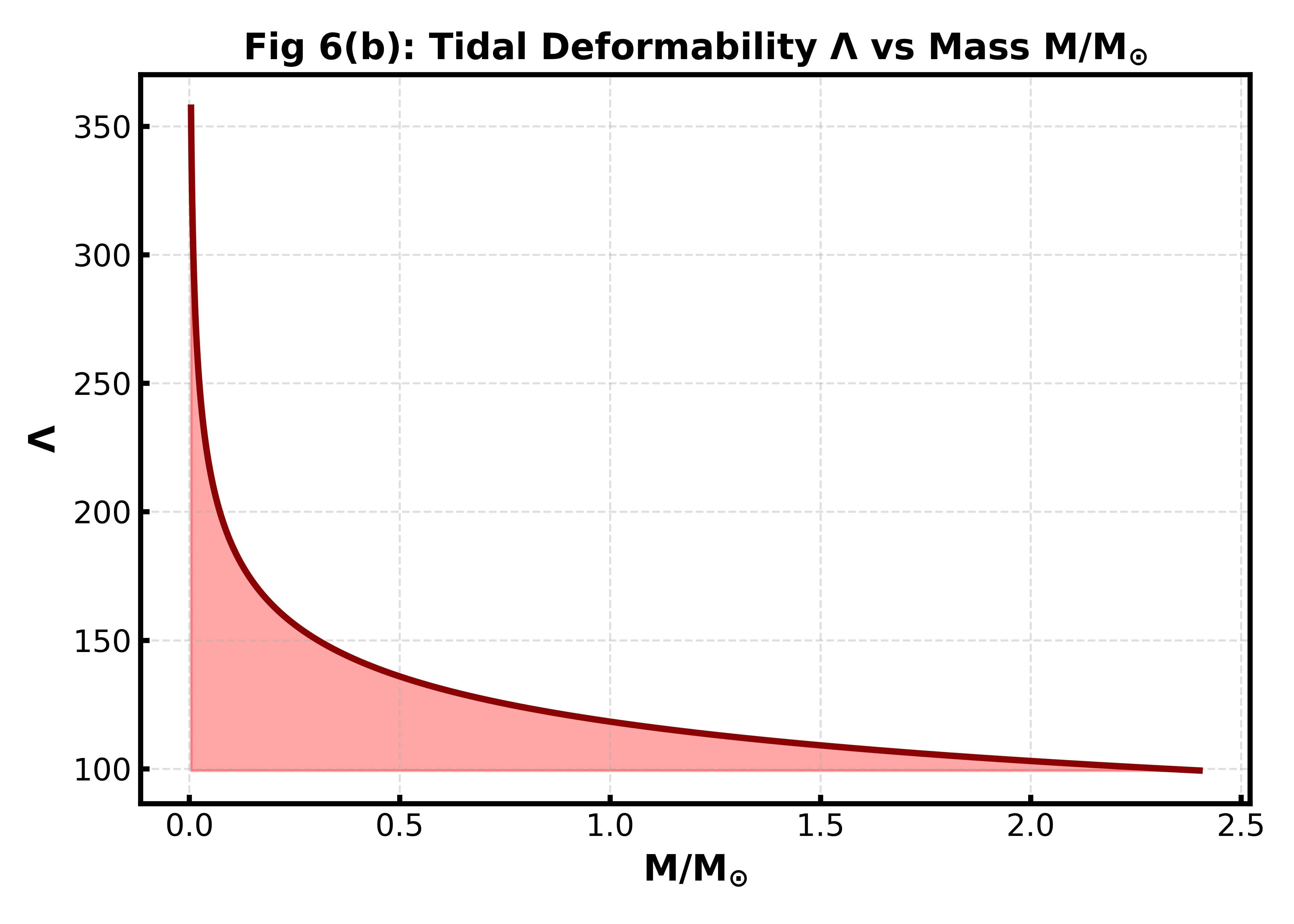}\\
\includegraphics[keepaspectratio=true,height=2.4in, width=3.5in]{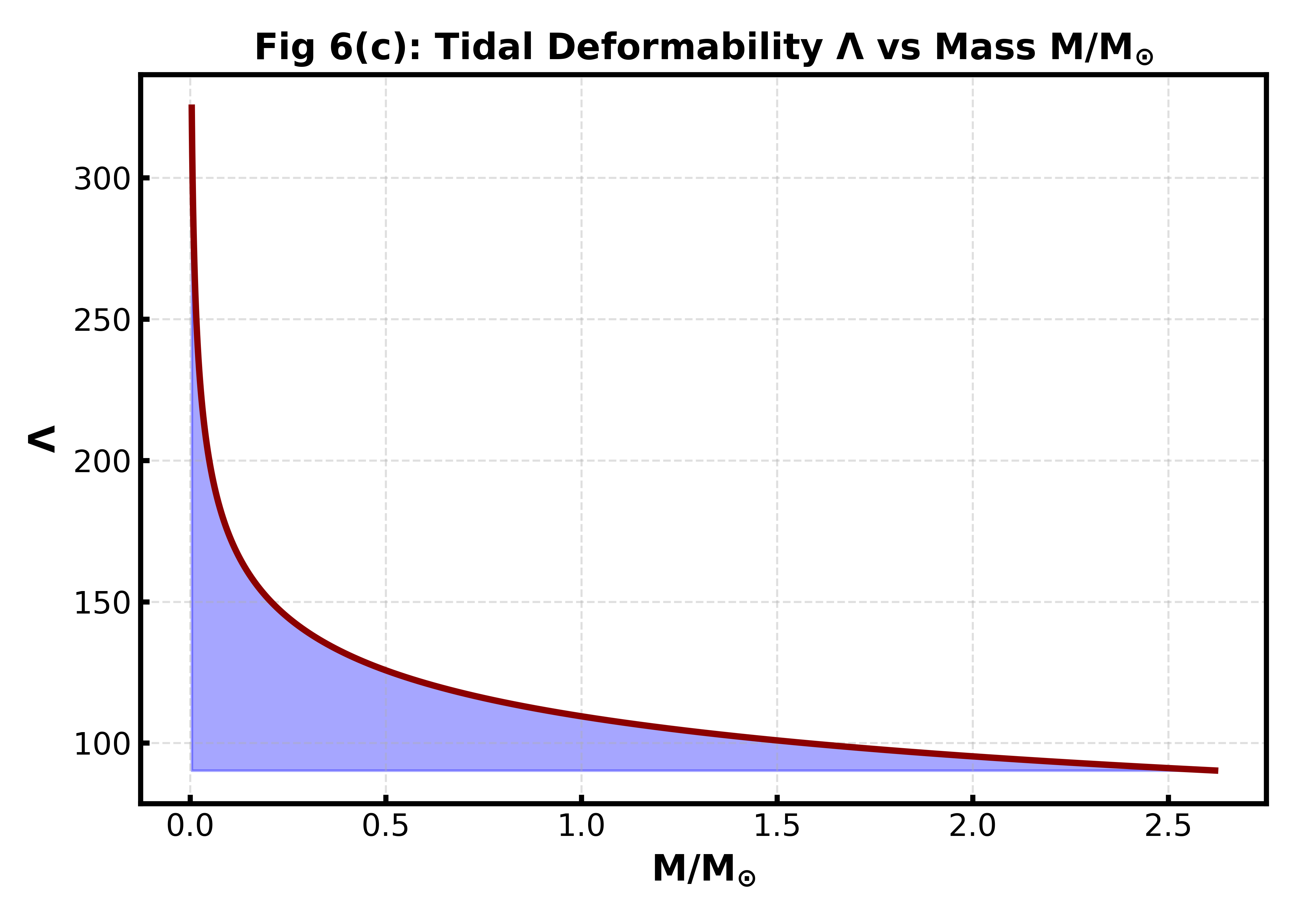}~~~~\includegraphics[keepaspectratio=true,height=2.4in, width=3.5in]{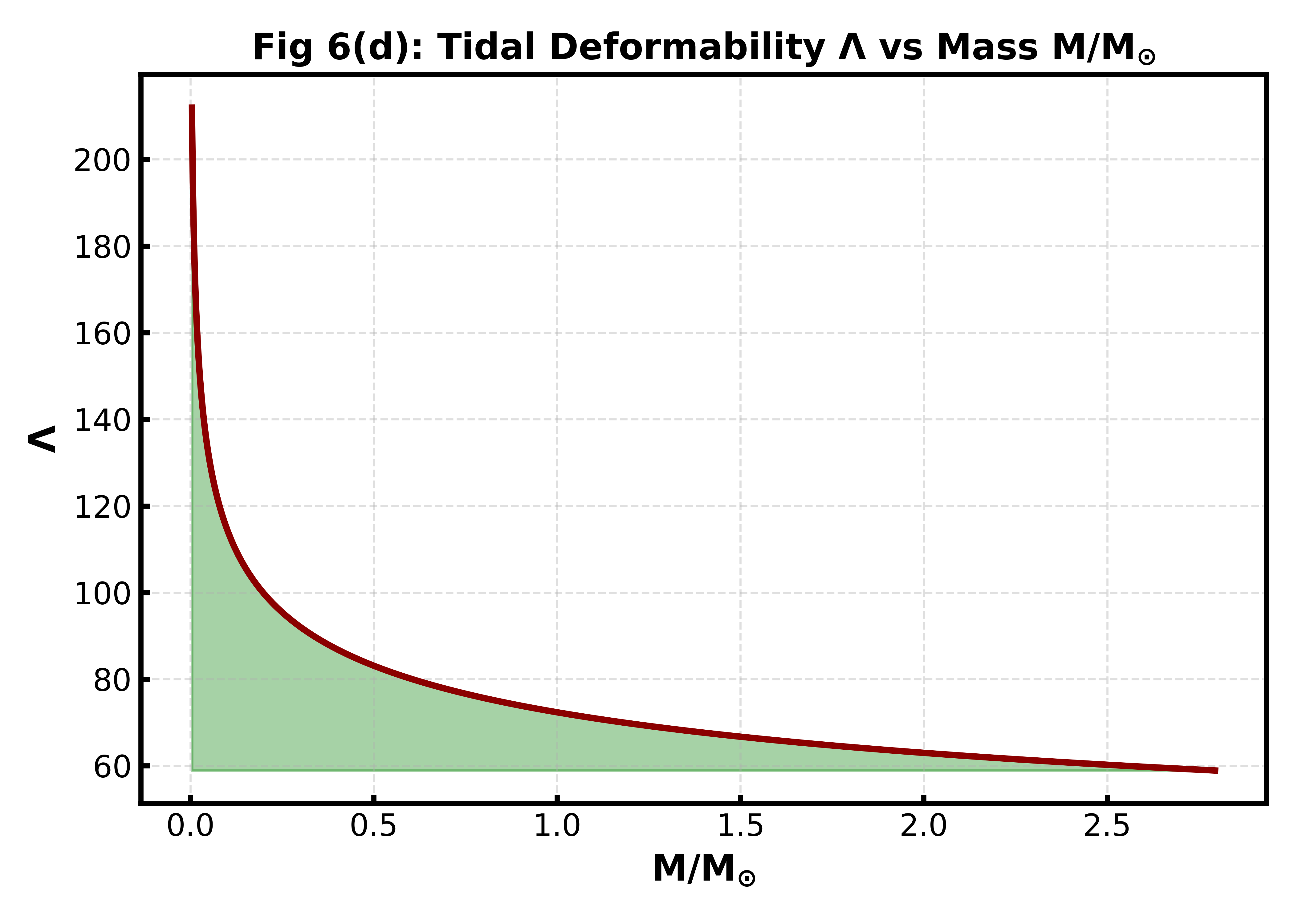}\\
\begin{justify}
Figures: 6(a--d) illustrate the variation of the dimensionless tidal deformability, $\Lambda$, as a function of the stellar mass, $M/M_{\odot}$, for the neutron stars (A) $PSR~J0437-4715$, (B) $PSR~J0614-3329$, (C) $Vela~X-1$, and (D) $PSR~J1614-2230$ within the framework of the constructed models. From the graphical profiles, it is evident that the dimensionless tidal deformability, $\Lambda$, decreases with increasing stellar mass, implying that more massive neutron stars are comparatively less deformable under an external tidal field. Furthermore, another interesting observation based on the graphical representation is that anisotropic stresses significantly suppress tidal deformability, particularly in the higher mass regime.\\
\end{justify}
\end{center}
\end{figure}


Here figures~6(a--d) present the variation of the dimensionless tidal deformability $\Lambda$ as a function of the stellar mass $M/M_{\odot}$ for the considered neutron-star configurations. The profiles clearly exhibit a monotonically decreasing behavior, indicating that $\Lambda$ decreases sharply with increasing stellar mass. This inverse correlation implies that low-mass neutron stars are more easily deformed by external tidal fields, whereas high-mass configurations are comparatively more compact and resistant to tidal deformation.

~~~A closer inspection of the curves reveals a steep decline in $\Lambda$ in the low-mass regime, followed by a gradual flattening at higher masses. This behavior reflects the increasing compactness of neutron stars with increasing mass, since the tidal deformability is highly sensitive to the compactness parameter $u=\frac{M(r)}{r}$. Physically, more compact stars possess stronger self-gravity, which suppresses tidal distortions. Conversely, low-mass neutron stars have smaller compactness and are therefore more susceptible to external tidal fields, resulting in comparatively larger values of $\Lambda$.

~~~An important feature emerging from these plots is that the obtained values of $\Lambda$ for intermediate- and high-mass neutron stars remain within the limits suggested by recent gravitational-wave observations. In particular, the binary neutron-star merger event GW170817 constrained the tidal deformability of a $1.4\,M_{\odot}$ neutron star to $\Lambda_{1.4}\lesssim800$ \cite{Abbott2017,Abbott2018}. The present model predicts values that are consistent with these observational constraints, thereby supporting its astrophysical viability.

~~~The systematic variation of the $\Lambda$--$M$ curves reflects the dependence of tidal deformability on the stellar structure predicted by the adopted equation of state. As the stellar mass increases, the compactness also increases, leading to a progressive reduction in the tidal response of the neutron star.

~~~An interesting feature of the present anisotropic model is that the predicted tidal deformability becomes comparatively smaller in the higher-mass regime. This behavior suggests that pressure anisotropy contributes to reducing the tidal response of compact stars, thereby providing an additional mechanism for supporting compact configurations while remaining compatible with current gravitational-wave observations.

~~~Furthermore, the smooth and non-intersecting nature of the $\Lambda$--$M$ curves demonstrates that the tidal deformability varies continuously with the stellar mass throughout the considered parameter space. The absence of any discontinuity or irregular behavior indicates that the proposed model remains mathematically regular and physically acceptable over the entire range of masses investigated.

~~~Therefore, the results presented in Figures~6(a--d) successfully reproduce the expected inverse relationship between the tidal deformability and the stellar mass. Overall, the obtained profiles demonstrate that the proposed anisotropic stellar model provides a physically consistent description of tidal deformability and remains compatible with current multi-messenger observational constraints.

\end{itemize}
\begin{figure}[H]
\begin{center}
\includegraphics[keepaspectratio=true,height=2.6in, width=3.5in]{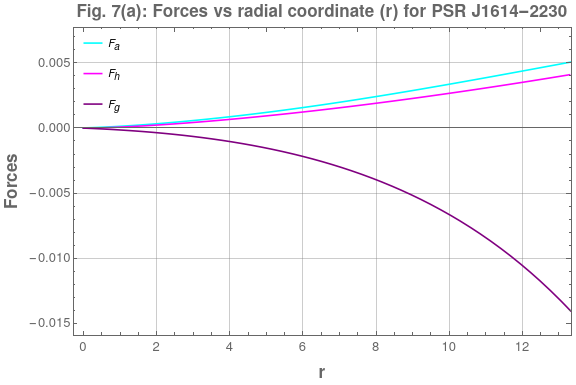}~~~~\includegraphics[keepaspectratio=true,height=2.6in, width=3.5in]{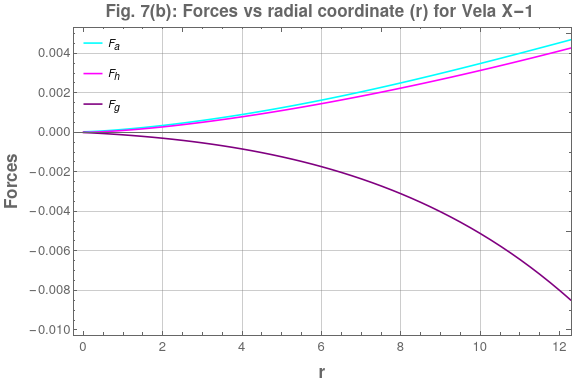}
\includegraphics[keepaspectratio=true,height=2.6in, width=3.5in]{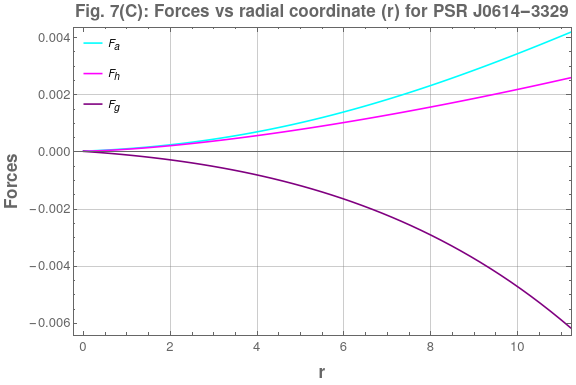}~~~~\includegraphics[keepaspectratio=true,height=2.6in, width=3.5in]{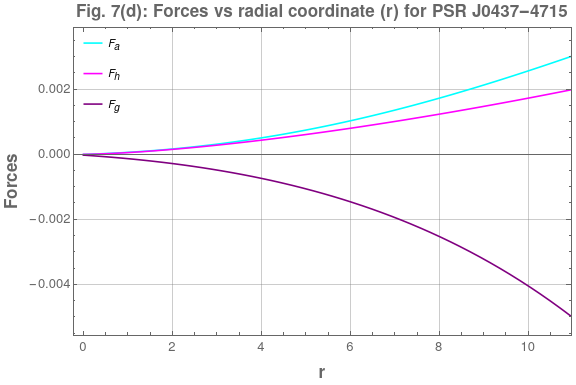}\\
\begin{justify}
Figures: 7(a--d) illustrate the variation of the different force profiles appearing in the generalized TOV equilibrium equation as functions of the radial coordinate, $r$, for the neutron stars (A) $PSR~J1614-2230$, (B) $Vela~X-1$, (C) $PSR~J0614-3329$, and (D) $PSR~J0437-4715$ within the framework of the constructed models. The graphical profiles demonstrate that the inward gravitational force ($F_g$) is effectively balanced by the combined action of the outward hydrostatic ($F_h$) and anisotropic ($F_a$) forces, thereby maintaining the hydrostatic equilibrium of the compact stellar configurations. Furthermore, the plots clearly illustrate the significant contribution of the anisotropic force to maintaining  the equilibrium configuration of the stellar model. Since the interior solution is physically meaningful only within the stellar boundary ($0 \leq r \leq R$), the profiles of the gravitational force ($F_g$), hydrostatic force ($F_h$) and anisotropic ($F_a$) force are presented only up to their respective stellar surfaces. This avoids extending the physical variables into the unphysical exterior region beyond the compact star.\\
\end{justify}
\end{center}
\end{figure}

\begin{itemize}

\item Figures~7(a--d) illustrate the radial variation of the three forces governing the hydrostatic equilibrium of the neutron-star configurations considered in the present work. The hydrostatic equilibrium of the stellar configuration is governed by the combined action of the gravitational force ($F_g$), the hydrostatic force ($F_h$), and the anisotropic force ($F_a$), whose mathematical expressions are given by \cite{Bowers1974,Herrera1997}
\begin{align}
F_g &= -\frac{1}{2}\left(\rho+p_r\right)\frac{dp_r}{dr},\qquad
F_h=-\frac{dp_r}{dr},\qquad
F_a=\frac{2}{r}\left(p_t-p_r\right).
\end{align}

~~~The graphical profiles clearly show that the gravitational force remains negative throughout the stellar interior, indicating its inward-directed nature, whereas both the hydrostatic and anisotropic forces remain positive and act outward against gravity. Throughout the stellar interior, these three forces satisfy the generalized Tolman--Oppenheimer--Volkoff (TOV) equilibrium condition,
\begin{equation}
F_g(r)+F_h(r)+F_a(r)=0,\qquad 0\leq r\leq R.
\end{equation}
The calculated force profiles demonstrate that this balance is maintained continuously from the stellar center to the stellar surface, thereby confirming that the proposed stellar configurations remain in complete hydrostatic equilibrium throughout the entire interior region.

~~~A closer examination of the profiles shows that the anisotropic force is negligibly small in the vicinity of the stellar center because
\[
\Delta(r)=p_t(r)-p_r(r)\rightarrow0
\qquad\text{as}\qquad
r\rightarrow0,
\]
as discussed in Figure~4. Consequently, the equilibrium near the center is primarily maintained through the balance between the gravitational and hydrostatic forces. As the radial coordinate increases, the contribution of the anisotropic force to the hydrostatic equilibrium becomes increasingly significant and provides additional outward support against the inward gravitational attraction.

~~~The smooth and continuous variation of all three forces indicates the absence of any discontinuity or irregular behavior within the stellar interior. Furthermore, the force profiles do not exhibit any abrupt change in magnitude or direction, demonstrating that the equilibrium configuration remains regular throughout the physically admissible region.

~~~The force profiles further indicate that the more massive neutron-star configurations exhibit comparatively larger gravitational, hydrostatic, and anisotropic forces than the lower-mass configurations. This behavior is expected because stronger gravitational fields require correspondingly larger pressure gradients and anisotropic stresses to maintain hydrostatic equilibrium.

~~~The present results are consistent with the generalized TOV equilibrium condition for anisotropic compact stars discussed by Bowers and Liang \cite{Bowers1974} and Herrera and Santos \cite{Herrera1997}, in which hydrostatic equilibrium is achieved through the combined balance of the gravitational, hydrostatic, and anisotropic forces.

~~~Overall, Figures~7(a--d) confirm that the proposed stellar model satisfies the generalized TOV equilibrium equation throughout the stellar interior. The combined action of the outward hydrostatic and anisotropic forces exactly balances the inward gravitational force, thereby establishing hydrostatic equilibrium and further demonstrating the physical acceptability of the constructed compact-star configurations.

\end{itemize}
\begin{itemize}

\item Figure~8 illustrates the radial variation of the gravitational redshift within the stellar interior, whose value at the stellar surface corresponds to the surface redshift $Z_s$. It is observed that the surface redshift remains bounded ($Z_s\lesssim3$), which lies well within the theoretical upper limit for anisotropic compact stars ($Z_s\le5$) \cite{Bowers1974,Ivanov2002}. Such moderate values are also consistent with current observational expectations, as extremely large redshifts have not been inferred for known neutron stars. Moreover, the more massive neutron-star configurations exhibit comparatively larger redshifts, reflecting their stronger gravitational fields and higher compactness.

~~~Figure~9 presents the radial variation of the compactness function,
\[
u(r)=\frac{M(r)}{r},
\]
which increases monotonically from the stellar center toward the surface. The compactness remains well below the Buchdahl limit ($u<4/9$) \cite{Buchdahl1959} throughout the stellar interior, confirming that the proposed stellar configurations satisfy one of the fundamental requirements for physically acceptable relativistic compact stars. Furthermore, the more massive neutron-star configurations attain comparatively larger compactness, in agreement with their stronger gravitational fields.

~~~Figure~10 illustrates the radial variation of the adiabatic index $\Gamma(r)$. It is evident that
\[
\Gamma>\frac{4}{3}
\]
throughout the stellar interior for all the neutron-star configurations considered, thereby satisfying the well-known dynamical stability criterion proposed by Chandrasekhar \cite{Chandrasekhar1964}. The gradual increase of the adiabatic index toward the stellar surface indicates an enhanced resistance of the stellar matter against radial compression. Consequently, the obtained profiles further support the dynamical stability and physical viability of the proposed compact-star model.

\end{itemize}

\begin{figure}[H]
\begin{center}
\includegraphics[keepaspectratio=true,height=2.6in, width=3.5in]{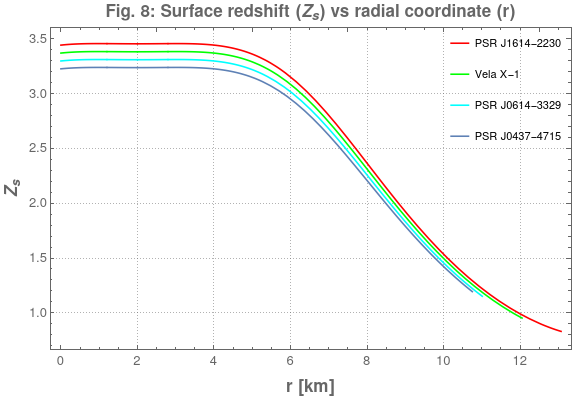}~~~~\includegraphics[keepaspectratio=true,height=2.6in, width=3.7in]{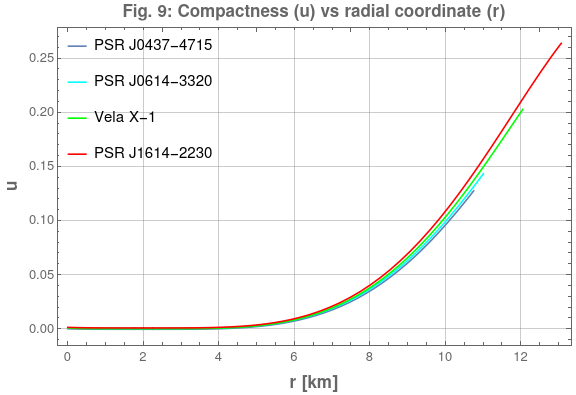}\\
\includegraphics[keepaspectratio=true,height=2.6in, width=3.5in]{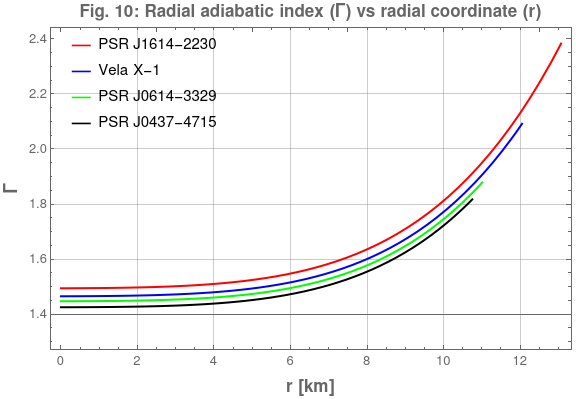}\\
\begin{justify}
Figures: 8--10 illustrate the profiles of the surface redshift, $Z_s$, compactness, $u(r)=\frac{M(r)}{r}$, and radial adiabatic index, $\Gamma(r)$, as functions of the radial coordinate, $r$, for the neutron stars (A) $PSR~J1614-2230$, (B) $Vela~X-1$, (C) $PSR~J0614-3329$, and (D) $PSR~J0437-4715$ within the framework of the constructed models. Since the interior solution is physically meaningful only within the stellar boundary ($0 \leq r \leq R$), the profiles of the surface redshift, $Z_s$, compactness, $u(r)=\frac{M(r)}{r}$, and radial adiabatic index, $\Gamma(r)$ are presented only up to their respective stellar surfaces. This avoids extending the physical variables into the unphysical exterior region beyond the compact star.\\
\end{justify}
\end{center}
\end{figure}
\begin{itemize}

\item Figures~11(a) and~11(b) illustrate the radial variation of the squares of the radial and tangential sound speeds,
\[
v_r^2=\frac{dp_r}{d\rho},
\qquad
v_t^2=\frac{dp_t}{d\rho},
\]
respectively, for the neutron-star configurations considered in the present work. It is evident from the graphical profiles that both sound speeds remain within the physically admissible interval
\[
0\leq v_r^2,\;v_t^2\leq1,
\]
throughout the stellar interior ($0\leq r\leq R$). Consequently, the proposed stellar model satisfies the causality condition, ensuring that the propagation speeds of pressure perturbations do not exceed the speed of light.

~~~Figure~11(a) shows that the squared radial sound speed remains nearly constant throughout the stellar interior, indicating a smooth and stable response of the radial pressure to variations in the energy density. In contrast, Figure~11(b) demonstrates that the squared tangential sound speed exhibits a radial variation while remaining continuous and free from any irregular behavior. The smooth nature of both profiles confirms the regularity of the interior solution.

~~~Furthermore, the difference between the squared sound speeds satisfies
\begin{equation}
-1\leq v_t^2-v_r^2\leq0,
\end{equation}
throughout the stellar interior, thereby fulfilling the Herrera cracking criterion \cite{Abreu2007}. This result indicates that the proposed anisotropic stellar configurations are potentially stable against local anisotropic perturbations.

~~~It is also observed that the sound-speed profiles corresponding to the four neutron-star configurations are remarkably similar, with only marginal quantitative differences. This behavior suggests that the adopted equation of state provides a consistent and physically acceptable description of matter over the range of stellar masses considered in the present analysis.

~~~Overall, the sound-speed profiles satisfy both the causality condition and the Herrera stability criterion, thereby providing additional evidence for the physical viability and local stability of the proposed compact-star model.
\end{itemize}
\begin{figure}[H]
\begin{center}
\includegraphics[keepaspectratio=true,height=2.6in, width=3.4in]{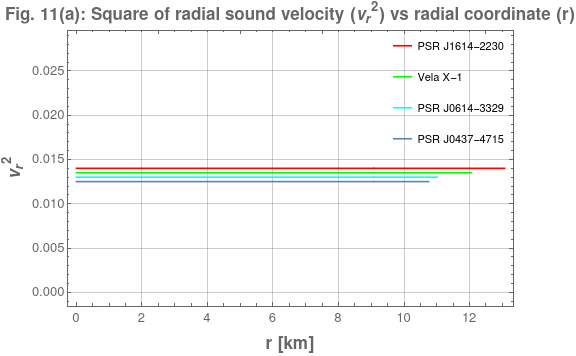}~~~~\includegraphics[keepaspectratio=true,height=2.5in, width=3.5in]{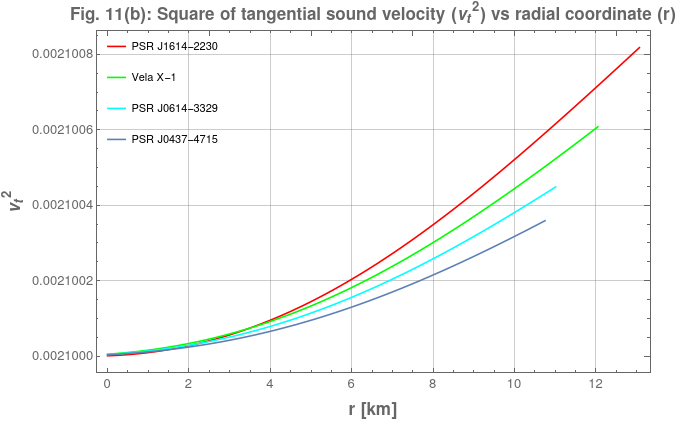}\\
\begin{justify}
Figures: 11(a--b) illustrate the variation of the squared radial sound speed, $v_{r}^{2}$, and the squared tangential sound speed, $v_{t}^{2}$, as functions of the radial coordinate, $r$, for the neutron stars (A) $PSR~J1614-2230$, (B) $Vela~X-1$, (C) $PSR~J0614-3329$, and (D) $PSR~J0437-4715$ within the framework of the constructed models. From the graphical representation it is clear that the difference between the squared sound speeds satisfies $-1 \leq v_t^2 - v_r^2 \leq 0 $ throughout the interior stellar configuration, which fulfills the Herrera cracking condition \cite{Abreu2007}. Since the interior solution is physically meaningful only within the stellar boundary ($0 \leq r \leq R$), the profiles of the radial sound speed, $v_{r}^{2}$, and the squared tangential sound speed, $v_{t}^{2}$, are presented only up to their respective stellar surfaces. This avoids extending the physical variables into the unphysical exterior region beyond the compact star.\\
\end{justify}
\end{center}
\end{figure}
\section{Observed Neutron Star Constraints and Physical Analysis}
To examine the physical acceptability and astrophysical applicability of the proposed stellar model, we consider a number of well-observed neutron stars whose observationally inferred masses and reported radii are listed in Table~\ref{tab:maxmassa} \cite{Linares2018,Cromartie2020,Abubekerov2008,Elebert2009,Demorest2010,Antoniadis2013,Rawls2011,Ozel2009,Romani2022,delaCruzDombriz2009,Kaaret1999,Freire2011,Guver2010a,Guver2010b}. For each neutron star, a representative mass is selected within its observationally constrained mass interval, and the corresponding stellar radius is determined from the mass--radius relation predicted by the present model. It is found that the estimated radius lies within the reported observational range, demonstrating good agreement between the theoretical predictions and the available astrophysical data.

Subsequently, using these mass--radius pairs, we evaluate the numerical values of the physical quantities required to characterize the stellar configurations. Furthermore, the maximum attainable mass and the corresponding stellar radius predicted by the proposed model are estimated for each neutron-star candidate. The resulting physical parameters are summarized in Tables~\ref{tab:maxmassb}--\ref{tab:maxmassd}.\begin{table}[H]
\small
\centering
\caption{\textbf{Observed mass--reported radius relation of various neutron stars and compact objects.}}
\label{tab:maxmassa}
\begin{center}
\vspace{0.1in}
\begin{tabular}{|p{0.6cm}|p{3.8cm}|p{2.0cm}|p{2.0cm}|p{1.1cm}|p{1.1cm}|p{1.5cm}|}
\hline
\multicolumn{7}{|c|}{Observed mass--reported radius relation of various neutron stars and compact objects}\\
\hline
Sl. No.&Neutron Star&$M_{Obs}(M_{\odot})$&$R_{Rep}[km]$&
$M[M_{\odot}]$&$R[km]$&$\frac{M(R)}{R}$\\
\hline
$L_{01}$&$Her~X-1$\cite{Rawls2011}&$0.85\pm0.15$&$7.20\pm0.41$&$0.85$&$7.20$&$0.174249$\\ \hline
$L_{02}$&$4U~1538-52$\cite{Ozel2009}&$0.87\pm0.07$&$7.86\pm0.21$&$0.86$&$7.86$&$0.161457$\\ \hline
$L_{03}$&$SAX~J1808.4-3658$\cite{Demorest2010}&$0.90\pm0.30$&$7.95\pm1.00$&$0.89$&$7.95$&$0.165239$\\ \hline
$L_{04}$&$LMC~X-4$\cite{Antoniadis2013}&$1.04\pm0.09$&$8.80\pm0.20$&$1.04$&$8.80$&$0.174439$\\ \hline
\rowcolor{gray!20}
$L_{05}$&$Cen~X-4$\cite{Maurya2023}&$1.14\pm0.45$&$9.75\pm 2.75$&$1.14$&$9.76$&$0.174436$\\ \hline
\rowcolor{gray!20}
$L_{06}$&$SMC~X-1$\cite{Maurya2024}&$1.21\pm0.12$&$11.50\pm 1.10$&$1.22$&$11.50$&$0.172402$\\ \hline
\rowcolor{gray!20}
$L_{07}$&$GW170817-2$\cite{Maurya2024}&$1.27\pm 0.09$&$11.90\pm 1.40$&$1.27$&$11.90$&$0.157523$\\ \hline
$L_{08}$&$SMC~X-4$\cite{Cromartie2020}&$1.29\pm0.05$&$9.95\pm0.09$&$1.29$&$9.95$&$0.191762$\\ \hline
$L_{09}$&$EXO~1785-248$\cite{Romani2022}&$1.30\pm0.20$&$10.83\pm0.44$&$1.30$&$10.83$&$0.177175$\\ \hline
\rowcolor{gray!20}
$L_{10}$&$PSR~J0030+0451$\cite{Maurya2025MassGap}&$1.34\pm0.16$&$12.70\pm0.80$&$1.33$&$12.70$&$0.152177$\\ \hline
\rowcolor{gray!20}
$L_{11}$&$X7~(47~Tuc)$\cite{Bogdanov2016}&$1.40\pm0.20$&$11.10\pm0.80$&$1.39$&$11.09$&$0.184999$\\ \hline
\rowcolor{gray!20}
$L_{12}$&$Swift~J1818.0-1607$\cite{Singh2026}&$1.40\pm 0.20$&$12.00\pm1.00$&$1.40$&$12.01$&$0.172057$\\ \hline
\rowcolor{gray!20}
$L_{13}$&$PSR~B0943+10$\cite{Maurya2026JCAP}&$1.40\pm 0.20$&$12.00\pm 1.00$&$1.41$&$12.00$&$0.173430$\\ \hline
\rowcolor{gray!20}
$L_{14}$&$M13~neutron~star$\cite{Shaw2018}&$1.40\pm0.30$&$12.30\pm 1.70$&$1.42$&$12.31$&$0.170262$\\ \hline
$L_{15}$&$Cen~X-3$\cite{delaCruzDombriz2009}&$1.49\pm0.08$&$10.17\pm0.13$&$1.48$&$10.17$&$0.214000$\\ \hline
\rowcolor{gray!20}
$L_{16}$&$PSR~B0656+14$\cite{Maurya2022JCAP}&$1.50\pm 0.20$&$12.50\pm 1.50$&$1.50$&$12.50$&$0.177120$\\ \hline
\rowcolor{gray!20}
$L_{17}$&$XTE~J1739-285$\cite{Kaaret2007}&$1.51\pm 0.20$&$10.90\pm 1.00$&$1.51$&$10.91$&$0.204286$\\ \hline
\rowcolor{gray!20}
$L_{18}$&$4U~1724-307$\cite{Hansraj2024}&$1.50\pm 0.40$&$12.00\pm0.50$&$1.53$&$11.98$&$0.188504$\\ \hline
$L_{19}$&$4U~1820-30$\cite{Kaaret1999} &$1.58\pm0.06$&$10.96\pm0.40$&$1.57$&$10.96$&$0.211432$\\ \hline
\rowcolor{gray!20}
$L_{20}$&$Aql~X-1$\cite{Li2017}&$1.60\pm 0.60$&$10.20\pm 0.24$&$1.60$&$10.20$&$0.231529$\\ \hline
\rowcolor{gray!20}
$L_{21}$&$KS~1731-260$\cite{Maurya2025FQT}&$1.61\pm 0.37$&$10.00\pm 2.20$&$1.62$&$10.01$&$0.238873$\\ \hline
$L_{22}$&$PSR~J1903+327$\cite{Linares2018}&$1.67\pm0.02$&$11.48\pm0.03$&$1.66$&$11.48$&$0.213225$\\ \hline
\rowcolor{gray!20}
$L_{23}$&$Cyg~X-2$\cite{Casares2010}&$1.71\pm0.21$&$11.50\pm 1.50$&$1.71$&$11.50$&$0.219475$\\ \hline
$L_{24}$&$4U~1608-52$\cite{Freire2011}&$1.74\pm0.01$&$11.30\pm0.10$&$1.74$&$11.30$&$0.227281$\\ \hline
$L_{25}$&$Vela~X-1$\cite{Abubekerov2008}&$1.77\pm0.08$&$11.75\pm0.85$&$1.77$&$12.56$&$0.208804$\\ \hline
$L_{26}$&$4U~1724-207$\cite{Guver2010a}&$1.81\pm0.27$&$12.20\pm1.40$&$1.80$&$12.20$&$0.218170$\\ \hline
$L_{27}$&$SAX~J1748.9-2021$\cite{Guver2010b}&$1.81\pm0.30$&$11.70\pm1.70$&$1.81$&$11.70$&$0.228332$\\ \hline
\rowcolor{gray!20}
$L_{28}$&$4U~1702-429$\cite{Singh2022Lovelock}&$1.90\pm0.30$&$12.40\pm0.40$&$1.90$&$12.39$&$0.226344$\\ \hline
$L_{29}$&$PSR~J1614-2230$\cite{Elebert2009}&$1.97\pm0.04$&$13.00\pm2.00$&$1.97$&$13.06$&$0.222643$\\ \hline
$L_{30}$&$PSR~J0348+0432$\cite{Abubekerov2008}&$2.01\pm0.04$&$13.00\pm2.00$&$2.01$&$13.00$&$0.228208$\\ \hline
$L_{31}$&$PSR~J0740+6620$\cite{Linares2018}&$2.08\pm0.17$&$12.39\pm1.50$&$2.08$&$12.39$&$0.247792$\\ \hline
$L_{32}$&$PSR~J1748-2446ad$\cite{Elebert2009}&$2.10\pm0.08$&$12.01\pm1.20$&$2.10$&$12.01$&$0.258077$\\ \hline
$L_{33}$&$PSR~J2215+5135$\cite{delaCruzDombriz2009}&$2.27\pm0.15$&$13.01\pm1.20$&$2.27$&$13.01$&$0.257533$\\ \hline
$L_{34}$&$PSR~J0952-0607$\cite{Rawls2011}&$2.35\pm0.17$&$10.01\pm1.50$&$2.35$&$10.01$&$0.346124$\\ \hline
\rowcolor{gray!20}
$L_{35}$&$4U~1700-37$\cite{Hansraj2025}&$2.44\pm0.27$&$11.50\pm 1.50$&$2.43$&$11.51$&$0.311614$\\ \hline
\rowcolor{gray!20}
$L_{36}$&$GW190814-2$\cite{Maurya2024ApJ}&$2.59\pm 0.09$&$11.00\pm 1.00$&$2.58$&$11.00$&$0.346189$\\ \hline
\end{tabular}
\end{center}
\vspace{2mm}
\footnotesize
\begin{justify}
\textit{\bf{Table~3:}} presents 36 compact-star data points for which the free parameters of the present model satisfy the condition $L>\tau r^{2}$, thereby ensuring a regular, singularity-free interior solution, as discussed in Eq.~\eqref{bal11}. The data are arranged in ascending order of stellar mass. The third and fourth columns list the reported (or observationally inferred) mass range, $M_{\rm Obs}/M_{\odot}$, and the corresponding reported radius, $R_{\rm Rep}$ (km), respectively. From each reported mass interval, a representative mass value, $M/M_{\odot}$, is selected (fifth column), and the corresponding attainable stellar radius $R$ (sixth column) is obtained from the theoretical mass--radius relation of the present model. The representative mass is chosen such that the resulting configuration simultaneously satisfies the regularity condition $L>\tau r^{2}$ and the boundary condition of the TOV equation, $p_r(R)=0$. The final column lists the ratio between the theoretically predicted mass and the corresponding reported radius. The entries labeled $L_{07}$ and $L_{36}$ correspond to the secondary components of two gravitational-wave events, for which the compactness-to-mass ratio is comparatively higher than that of the remaining objects in the sample.
\end{justify}
\end{table}
\noindent
\begin{itemize}

\item In continuation of the above discussion, Table~\ref{tab:maxmassa} presents the observationally inferred masses and reported radii of several well-known neutron stars and compact objects together with the corresponding theoretical estimates obtained from the proposed stellar model. It is evident that the predicted masses ($M$) and corresponding radii ($R$) agree well with the observed masses ($M_{\mathrm{Obs}}$) and reported radii ($R_{\mathrm{Rep}}$) within their respective uncertainty ranges. This agreement demonstrates the observational consistency and astrophysical applicability of the proposed stellar model.

~~~A careful examination of the compactness parameter,
\[
u=\frac{M}{R},
\]
shows that all stellar configurations remain well below the Buchdahl limit, thereby supporting the earlier conclusion drawn from Figure~9 that the proposed model describes physically acceptable compact stars. Moreover, even for massive pulsars such as PSR~J1614$-$2230, PSR~J0348$+$0432, and PSR~J0740$+$6620, the predicted radii remain compatible with the ranges inferred from recent observational studies, including the NICER mission.

~~~Another important outcome is that the proposed model successfully accommodates neutron stars over a broad mass interval ($\sim0.8$--$2.3\,M_{\odot}$), demonstrating its applicability to both low- and high-mass compact stars. Furthermore, the estimated maximum masses obtained from the model are systematically higher than the corresponding observed masses, indicating that the proposed framework is capable of supporting more massive equilibrium configurations.

~~~The good agreement between the predicted and observed mass--radius pairs provides additional support for the physical acceptability of the constructed stellar model. In particular, the obtained stellar radii are consistent with the monotonic density and pressure distributions presented in Figures~1--3, while the corresponding compactness values remain compatible with the gravitational redshift profiles discussed in Figure~8. Together, these results indicate that the proposed model consistently satisfies the fundamental physical requirements expected for realistic compact stars.

~~~It is also observed that the proposed anisotropic model is capable of describing both comparatively low-mass and massive neutron stars within a single theoretical framework. The additional outward support provided by the positive anisotropy, together with the hydrostatic pressure gradient, contributes to maintaining equilibrium against gravitational collapse, as demonstrated in Figures~4 and~7.

~~~Overall, the results presented in Table~\ref{tab:maxmassa} provide strong observational support for the proposed stellar model. The close agreement between the theoretical predictions and the available observational data, together with the fulfillment of the equilibrium, compactness, causality, and stability criteria established in the preceding sections, demonstrates that the present framework provides a physically consistent and observationally viable description of relativistic compact stars.
\end{itemize}
\begin{itemize}

\item Tables~\ref{tab:maxmassb}--\ref{tab:maxmassd} provide a comprehensive quantitative verification of the graphical analyses presented in the preceding sections. While Figures~1--11 establish the physical behavior of the proposed stellar model through continuous radial profiles, the tabulated numerical values confirm these results for a representative sample of 36 compact stars spanning a broad range of observed masses.

~~~Table~\ref{tab:maxmassb} lists the model parameters ($L$, $\tau$, $S$), together with the estimated maximum mass ($M_{\mathrm{Est}}$) and the corresponding stellar radius ($R_{\mathrm{Est}}$). It is observed that the model parameters vary smoothly and systematically across the stellar sample, indicating a well-defined and self-consistent parameter space. Furthermore, the estimated maximum masses increase monotonically with the stellar configurations while the corresponding radii remain within the range expected for realistic neutron stars. These numerical results quantitatively support the mass--radius characteristics discussed in Figure~5 and demonstrate that the proposed model is capable of describing massive compact stars without violating the fundamental compactness constraints.

~~~The numerical values presented in Table~\ref{tab:maxmassc} further confirm the physical properties discussed in Figures~1--4. In particular, the inequalities
\[
\rho_c>\rho_s
\qquad\text{and}\qquad
p_t>p_r
\]
are satisfied for every compact-star configuration considered in this study. The former confirms the monotonic decrease of the matter density from the stellar centre to the surface, whereas the latter verifies the existence of positive pressure anisotropy throughout the stellar interior. In addition, both the pressure components and the matter density exhibit a gradual increase with stellar mass, reflecting the stronger internal pressure support required to maintain equilibrium in more massive neutron-star configurations.

~~~Table~\ref{tab:maxmassd} provides a quantitative confirmation of the stability and compactness analyses presented in Figures~8--11. The radial adiabatic index satisfies
\[
2.1385\le\Gamma\le2.4529,
\]
which is well above the Chandrasekhar stability criterion ($\Gamma>4/3$) for all stellar models, thereby confirming their dynamical stability against infinitesimal radial perturbations. Similarly, the surface redshift increases gradually from
\[
0.9283\le Z_s\le1.1407,
\]
while remaining well below the theoretical upper bound ($Z_s\le5$) for anisotropic compact stars. Moreover, the estimated compactness parameter satisfies
\[
0.2206\le
\frac{M_{\mathrm{Est}}}{R_{\mathrm{Est}}}
\le0.3954,
\]
which remains comfortably below the Buchdahl limit
\[
\frac{M}{R}<\frac49.
\]
These numerical results are in excellent agreement with the corresponding graphical profiles and demonstrate that the proposed stellar configurations satisfy the accepted compactness and gravitational-redshift constraints.

~~~Taken together, Tables~\ref{tab:maxmassb}--\ref{tab:maxmassd} quantitatively confirm that the proposed stellar model consistently satisfies all the essential physical requirements expected for realistic relativistic compact stars. The numerical values validate the regular density and pressure distributions, positive pressure anisotropy, generalized TOV equilibrium, dynamical stability, causality, acceptable compactness, and moderate surface redshift established in the preceding analyses. Furthermore, the close agreement between the theoretical predictions and the available observational data demonstrates the robustness, mathematical consistency, and astrophysical viability of the proposed anisotropic stellar model within the framework of Einstein's gravity.
\end{itemize}
\begin{table}[H]
\small
\centering
\caption{\textbf{model constants and estimated maximum mass of various neutron stars for constructed stellar model}}
\label{tab:maxmassb}
\begin{center}
\vspace{0.1 in}
\begin{tabular}{
|p{0.4cm}|p{3.8cm}|p{1.3cm}|p{1.3cm}|p{1.2cm}|p{1.6cm}|p{1.5cm}|}
\hline
\multicolumn{7}{|c|}
{model constants for the neutron stars .}\\
\hline
Sl. No.&Neutron Star& $L[km^{0}]$&$\tau [km^{-2}]$& $S [km^{0}]$& $M_{Est}(M_{\odot})$&$R_{Est}[km]$\\
\hline
$L_{01}$&$Her~X-1$\cite{Rawls2011}& $5.5609$&$0.0179$& $0.5381$& $1.5209$ & $10.1775$\\
\hline
$L_{02}$&$4U~1538-52$\cite{Ozel2009}& $5.6347$&$0.0231$& $0.5539$& $1.8207$ & $10.3257$\\
\hline
$L_{03}$&$SAX~J1808.4-3658$\cite{Demorest2010}& $5.7795$&$0.0258$& $0.5904$& $1.8779$ & $10.7135$\\
\hline
$L_{04}$&$LMC~X-4$\cite{Antoniadis2013}& $6.6255$ &$0.0271$& $0.6140$ &$1.9085$ & $10.8293$\\
\hline
\rowcolor{gray!20}
$L_{05}$&$Cen~X-4$\cite{Maurya2023}&$6.6731$&$0.0279$&$0.6185$&$1.9273$&$10.8341$\\ \hline
\rowcolor{gray!20}
$L_{06}$&$SMC~X-1$\cite{Maurya2024}&$6.7028$&$0.0281$&$0.6209$&$1.9405$&$11.1427$\\ \hline
\rowcolor{gray!20}
$L_{07}$&$GW170817-2$\cite{Maurya2024}&$6.7225$&$0.0285$&$0.6285$&$1.9711$&$11.2713$\\ \hline
$L_{08}$&$SMC~X-4$\cite{Cromartie2020}& $6.7416$&$0.0286$& $0.6391$& $2.0351$ & $10.8628$\\
\hline
$L_{09}$&$EXO~1785-248$\cite{Romani2022}& $6.7603$&$0.0319$& $0.6507$& $2.2362$ & $11.2379$\\
\hline
\rowcolor{gray!20}
$L_{10}$&$PSR~J0030+0451$\cite{Maurya2025MassGap}&$6.7629$&$0.0321$&$0.6516$&$2.2943$&$11.7481$\\ \hline
\rowcolor{gray!20}
$L_{11}$&$X7~(47~Tuc)$\cite{Bogdanov2016}&$6.7673$&$0.0340$&$0.6547$&$2.3109$&$11.1917$\\ \hline
\rowcolor{gray!20}
$L_{12}$&$Swift~J1818.0-1607$\cite{Singh2026}&$6.7682$&$0.0352$&$0.6571$&$2.3427$&$11.4095$\\ \hline
\rowcolor{gray!20}
$L_{13}$&$PSR~B0943+10$\cite{Maurya2026JCAP}&$6.7730$&$0.0357$&$0.6583$&$2.3502$&$11.3754$\\ \hline
\rowcolor{gray!20}
$L_{14}$&$M13~neutron~star$\cite{Shaw2018}&$6.7783$&$0.0364$&$0.6592$&$2.3685$&$11.5273$\\ \hline
$L_{15}$&$Cen~X-3$\cite{delaCruzDombriz2009}& $6.7891$&$0.0373$& $0.6619$& $2.4385$ & $10.7544$\\
\hline
\rowcolor{gray!20}
$L_{16}$&$PSR~B0656+14$\cite{Maurya2022JCAP}&$6.8157$&$0.0392$&$0.6653$&$2.4425$&$11.6328$\\ \hline
\rowcolor{gray!20}
$L_{17}$&$XTE~J1739-285$\cite{Kaaret2007}&$6.8648$&$0.0407$&$0.6685$&$2.4522$&$10.9425$\\ \hline
\rowcolor{gray!20}
$L_{18}$&$4U~1724-307$\cite{Hansraj2024}&$6.8825$&$0.0416$&$0.6741$&$2.4594$&$11.5178$\\ \hline
$L_{19}$&$4U~1820-30$\cite{Kaaret1999}  & $6.9782$&$0.0429$& $0.6794$& $2.4792$ & $11.1720$\\
\hline
\rowcolor{gray!20}
$L_{20}$&$Aql~X-1$\cite{Li2017}&$7.2806$&$0.0433$&$0.6833$&$2.4807$&$10.7743$\\ \hline
\rowcolor{gray!20}
$L_{21}$&$KS~1731-260$\cite{Maurya2025FQT}&$7.5113$&$0.0437$&$0.6956$&$2.4916$&$10.5283$\\ \hline
$L_{22}$&$PSR~J1903+327$\cite{Linares2018}& $7.6157$&$0.0440$& $0.7107$& $2.5117$ & $11.7972$\\
\hline
\rowcolor{gray!20}
$L_{23}$&$Cyg~X-2$\cite{Casares2010}&$7.7853$&$0.0459$&$0.7129$&$2.5538$&$11.8117$\\ \hline
$L_{24}$&$4U~1608-52$\cite{Freire2011}& $7.9159$&$0.0468$& $0.7136$& $2.5711$ & $11.6105$\\
\hline
$L_{25}$&$Vela~X-1$\cite{Abubekerov2008}& $8.5261$&$0.0481$& $0.7204$& $2.6074$ & $12.0518$\\
\hline
$L_{26}$& $4U~1724-207$\cite{Guver2010a}      & $8.9374$&$0.0487$& $0.7261$& $2.6341$ & $12.1249$\\
\hline
$L_{27}$ & $SAX~J1748.9-2021$\cite{Guver2010b} & $9.4759$&$0.0493$& $0.7344$& $2.7038$ & $11.8571$\\
\hline
\rowcolor{gray!20}
$L_{28}$&$4U~1702-429$\cite{Singh2022Lovelock}&$9.6078$&$0.0495$&$0.7371$&$2.7328$&$12.1513$\\ \hline
$L_{29}$ & $PSR~J1614-2230$\cite{Elebert2009}   & $9.7173$&$0.0497$& $0.7502$& $2.7645$ & $13.0729$\\
\hline
$L_{30}$&  $PSR~J0348+0432$\cite{Abubekerov2008}     & $9.8338$&$0.0514$& $0.7581$& $2.7735$ & $13.1705$\\
\hline
$L_{31}$&  $PSR~J0740+6620$\cite{Linares2018}     & $9.8790$&$0.0523$& $0.7607$& $2.8194$ & $12.6206$\\
\hline
$L_{32}$&  $PSR~J1748-2446ad$\cite{Elebert2009} & $9.9227$&$0.0528$& $0.7651$& $2.8605$ & $12.1731$\\
\hline
$L_{33}$ & $PSR~J2215+5135$ \cite{delaCruzDombriz2009}      & $9.9876$&$0.0533$& $0.7739$& $2.9244$ & $13.2131$\\
\hline
$L_{34}$ & $PSR~J0952-0607$\cite{Rawls2011}   & $10.0275$&$0.0547$& $0.7884$& $2.9518$ & $10.9338$\\
\hline
\rowcolor{gray!20}
$L_{35}$&$4U~1700-37$\cite{Hansraj2025}&$10.1518$&$0.0552$&$0.7895$&$2.9637$&$11.8124$\\ \hline
\rowcolor{gray!20}
$L_{36}$&$GW190814-2$\cite{Maurya2024ApJ}&$10.1703$&$0.0583$&$0.7904$&$2.9829$&$11.1326$\\ \hline
\end{tabular}
\end{center}
\vspace{2mm}
\footnotesize
\begin{justify}
\textit{\bf{Table~4:}} presents the model parameters corresponding to the 36 compact stars considered in the present study. The third, fourth, and fifth columns list the values of the model parameters $L$, $\tau$, and $S$, respectively, from which it is evident that all stellar configurations satisfy the regularity condition $L>\tau r^{2}$. Whereas, the sixth and seventh columns respectively provide the maximum estimated mass ($M_{\rm Est}$) of each compact star, together with the corresponding stellar radius ($R_{\rm Est}$) at which the maximum mass is attained. These results demonstrate that all the considered stellar models satisfy the imposed physical constraints while yielding well-defined maximum mass and radius configurations.
\end{justify}
\end{table}
\begin{table}[H]
\small
\centering
\caption{\textbf{Radial pressure($p_r$), Tangential pressure($p_t$), surface density($\rho_s$) and Central density($\rho_c$) of various neutron stars for constructed stellar model}}
\label{tab:maxmassc}
\begin{center}
\vspace{0.1 in}
\begin{tabular}{
|p{1.2cm}|p{3.8cm}|p{2.4cm}|p{2.4cm}|p{2.4cm}|p{2.4cm}|}
\hline
\multicolumn{6}{|c|}
{Radial pressure($p_r$), Tangential pressure($p_t$), surface density($\rho_s$) and Central density($\rho_c$)}\\
\hline
Sl. No.&Compact Star& $p_r$($N/m^{2}$)&$p_t$($N/m^{2}$)& $\rho_s|_{r=R}$($kg/m^{3}$)& $\rho_c|_{r=0}$($kg/m^{3}$)\\
\hline
$L_{01}$ & $Her~X-1$\cite{Rawls2011} & $3.7593\times 10^{34}$ & $6.5074\times 10^{34}$ & $1.0863\times 10^{17}$ & $1.7349\times 10^{18}$ \\
\hline
$L_{02}$ & $4U~1538-52$\cite{Ozel2009} & $3.9047\times 10^{34}$ & $6.6285\times 10^{34}$ & $1.1585\times 10^{17}$ & $1.9021\times 10^{18}$ \\
\hline
$L_{03}$ & $SAX~J1808.4-3658$\cite{Demorest2010} & $4.1164\times 10^{34}$ & $6.8193\times 10^{34}$ & $1.1937\times 10^{17}$ & $2.2150\times 10^{18}$ \\
\hline
$L_{04}$&$LMC~X-4$\cite{Antoniadis2013}& $4.4092\times 10^{34}$&$7.1427\times 10^{34}$& $2.1152\times 10^{17}$& $2.2793\times 10^{18}$\\ \hline
\rowcolor{gray!20}
$L_{05}$ & $Cen~X-4$\cite{Maurya2023} & $4.4307\times 10^{34}$ & $7.1507\times 10^{34}$ & $2.1241\times 10^{17}$ & $2.2944\times 10^{18}$\\ \hline
\rowcolor{gray!20}
$L_{06}$ & $SMC~X-1$\cite{Maurya2024} & $4.4591\times 10^{34}$ & $7.1569\times 10^{34}$ & $2.1308\times 10^{17}$ &$2.2970\times 10^{18}$\\ \hline
\rowcolor{gray!20}
$L_{07}$ & $GW170817-2$\cite{Maurya2024} & $4.5181\times 10^{34}$ & $7.1688\times 10^{34}$ & $2.1455\times 10^{17}$ & $2.3086\times 10^{18}$\\ \hline
$L_{08}$&$SMC~X-4$\cite{Cromartie2020}& $4.5273\times 10^{34}$&$7.1709\times 10^{34}$& $2.1471\times 10^{17}$& $2.3379\times 10^{18}$ \\
\hline
$L_{09}$&$EXO~1785-248$\cite{Romani2022}& $4.9305\times 10^{34}$&$7.3866\times 10^{34}$& $2.2106\times 10^{17}$& $2.3701\times 10^{18}$ \\
\hline
\rowcolor{gray!20}
$L_{10}$&$PSR~J0030+0451$\cite{Maurya2025MassGap}& $4.9722\times 10^{34}$ & $7.3885\times 10^{34}$ & $2.2193\times 10^{17}$ & $2.3774\times 10^{18}$\\ \hline
\rowcolor{gray!20}
$L_{11}$&$X7~(47~Tuc)$\cite{Bogdanov2016}& $5.1930\times 10^{34}$ & $7.4317\times 10^{34}$ & $2.2376\times 10^{17}$ & $2.4117\times 10^{18}$\\ \hline
\rowcolor{gray!20}
$L_{12}$&$Swift~J1818.0-1607$\cite{Singh2026}& $5.2293\times 10^{34}$ & $7.4592\times 10^{34}$ & $2.2418\times 10^{17}$ & $2.4695\times 10^{18}$\\ \hline
\rowcolor{gray!20}
$L_{13}$&$PSR~B0943+10$\cite{Maurya2026JCAP}& $5.3150\times 10^{34}$ & $7.4690\times 10^{34}$ & $2.2482\times 10^{17}$ & $2.4803\times 10^{18}$\\ \hline
\rowcolor{gray!20}
$L_{14}$&$M13~neutron~star$\cite{Shaw2018}& $5.3865\times 10^{34}$ & $7.4971\times 10^{34}$ & $2.2517\times 10^{17}$ & $2.4916\times 10^{18}$\\ \hline
$L_{15}$&$Cen~X-3$\cite{delaCruzDombriz2009}& $5.5910\times 10^{34}$&$7.5159\times 10^{34}$& $2.2609\times 10^{17}$& $2.5038\times 10^{18}$ \\
\hline
\rowcolor{gray!20}
$L_{16}$&$PSR~B0656+14$\cite{Maurya2022JCAP}& $5.6368\times 10^{34}$ & $7.5758\times 10^{34}$ & $2.2665\times 10^{17}$ & $2.5289\times 10^{18}$\\ \hline
\rowcolor{gray!20}
$L_{17}$&$XTE~J1739-285$\cite{Kaaret2007}& $5.6805\times 10^{34}$ & $7.6855\times 10^{34}$ & $2.2849\times 10^{17}$ & $2.5516\times 10^{18}$\\ \hline
\rowcolor{gray!20}
$L_{18}$&$4U~1724-307$\cite{Hansraj2024}& $5.7329\times 10^{34}$ & $7.7094\times 10^{34}$ & $2.3118\times 10^{17}$ & $2.5755\times 10^{18}$\\ \hline
$L_{19}$&$4U~1820-30$\cite{Kaaret1999}  & $5.8114\times 10^{34}$&$7.7822\times 10^{34}$& $2.3327\times 10^{17}$& $2.5975\times 10^{18}$ \\
\hline
\rowcolor{gray!20}
$L_{20}$&$Aql~X-1$\cite{Li2017}& $5.8753\times 10^{34}$ & $8.0983\times 10^{34}$ & $2.3975\times 10^{17}$ & $2.6495\times 10^{18}$\\ \hline
\rowcolor{gray!20}
$L_{21}$&$KS~1731-260$\cite{Maurya2025FQT}& $5.9374\times 10^{34}$ & $8.2295\times 10^{34}$ & $2.4752\times 10^{17}$ & $2.7570\times 10^{18}$\\ \hline
$L_{22}$&$PSR~J1903+327$\cite{Linares2018}& $6.3958\times 10^{34}$&$8.5081\times 10^{34}$& $2.5708\times 10^{17}$& $2.8492\times 10^{18}$ \\
\hline
\rowcolor{gray!20}
$L_{23}$&$Cyg~X-2$\cite{Casares2010}& $6.4659\times 10^{34}$ & $8.7440\times 10^{34}$ & $2.6583\times 10^{17}$ & $3.0937\times 10^{18}$\\ \hline
$L_{24}$&$4U~1608-52$\cite{Freire2011}& $6.8691\times 10^{34}$&$8.9423\times 10^{34}$& $2.8022\times 10^{17}$& $3.1095\times 10^{18}$ \\
\hline
$L_{25}$&$Vela~X-1$\cite{Abubekerov2008}& $8.0266\times 10^{34}$&$9.8762\times 10^{34}$& $3.1108\times 10^{17}$& $3.4528\times 10^{18}$ \\
\hline
$L_{26}$& $4U~1724-207$\cite{Guver2010a}      & $8.1094\times 10^{34}$&$9.8906\times 10^{34}$& $3.1753\times 10^{17}$& $3.4629\times 10^{18}$ \\
\hline
$L_{27}$ & $SAX~J1748.9-2021$\cite{Guver2010b} & $8.1103\times 10^{34}$&$9.9177\times 10^{34}$& $3.2306\times 10^{17}$& $3.4707\times 10^{18}$ \\
\hline
\rowcolor{gray!20}
$L_{28}$&$4U~1702-429$\cite{Singh2022Lovelock}& $8.1770\times 10^{34}$ & $9.9205\times 10^{34}$ & $3.2391\times 10^{17}$ & $3.4728\times 10^{18}$\\ \hline
$L_{29}$ & $PSR~J1614-2230$ \cite{Elebert2009}  & $8.1375\times 10^{34}$&$9.9374\times 10^{34}$& $3.2579\times 10^{17}$& $3.4736\times 10^{18}$ \\
\hline
$L_{30}$&  $PSR~J0348+0432$\cite{Abubekerov2008}     & $8.2841\times 10^{34}$&$9.9591\times 10^{34}$& $3.3479\times 10^{17}$& $3.4835\times 10^{18}$ \\
\hline
$L_{31}$&  $PSR~J0740+6620$\cite{Linares2018}     & $8.3575\times 10^{34}$&$1.0347\times 10^{35}$& $3.3505\times 10^{17}$& $3.5073\times 10^{18}$ \\
\hline
$L_{32}$&  $PSR~J1748-2446ad$\cite{Elebert2009} & $8.4309\times 10^{34}$&$1.1251\times 10^{35}$& $3.3621\times 10^{17}$& $3.5480\times 10^{18}$ \\
\hline
$L_{33}$ & $PSR~J2215+5135$ \cite{delaCruzDombriz2009}      & $8.4729\times 10^{34}$&$1.2508\times 10^{35}$& $3.7095\times 10^{17}$& $3.6003\times 10^{18}$ \\
\hline
$L_{34}$ & $PSR~J0952-0607$\cite{Rawls2011}   & $8.4922\times 10^{34}$&$1.3173\times 10^{35}$& $3.7330\times 10^{17}$& $3.6258\times 10^{18}$ \\
\hline
\rowcolor{gray!20}
$L_{35}$&$4U~1700-37$\cite{Hansraj2025}& $8.5407\times 10^{34}$ & $1.5739\times 10^{35}$ & $3.7447\times 10^{17}$ & $3.6866\times 10^{18}$\\ \hline
\rowcolor{gray!20}
$L_{36}$&$GW190814-2$\cite{Maurya2024ApJ}& $8.5626\times 10^{34}$ & $1.6174\times 10^{35}$ & $3.7530\times 10^{17}$ & $3.6915\times 10^{18}$\\ \hline
\end{tabular}
\end{center}
\vspace{2mm}
\footnotesize
\begin{justify}
\textit{\bf{Table~5:}} summarizes the numerical values of the principal physical quantities predicted by the present stellar model for the 36 compact stars considered in this study. Specifically, the table lists the radial pressure ($p_r$), tangential pressure ($p_t$), and matter density profile ($\rho$) evaluated at both the stellar centre ($r=0$) and the stellar surface ($r=R$). It is evident from the tabulated results that all the physical parameters remain within the expected physically admissible range for realistic neutron-star configurations. Furthermore, the pressure and matter density exhibit an overall increasing trend with increasing stellar mass. In addition, the inequality $p_t>p_r$ is satisfied for all stellar configurations considered, indicating the existence of positive pressure anisotropy throughout the stellar interior.
\end{justify}
\end{table}
\begin{table}[H]
\small
\centering
\caption{\textbf{Adiabatic index[$\Gamma$], Surface redshift($Z_s$) and estimated compactness($\frac{M_{Est}}{R_{Est}}$) of various neutron stars for constructed stellar model}}
\label{tab:maxmassd}
\begin{center}
\vspace{0.1 in}
\begin{tabular}{
|p{1.2cm}|p{3.8cm}|p{1.5cm}|p{1.5cm}|p{1.5cm}|}
\hline
\multicolumn{5}{|c|}
{Adiabatic index[$\Gamma$], Surface redshift($Z_s$) and estimated compactness($\frac{M_{Est}}{R_{Est}}$) }\\
\hline
Sl. No.&Compact Star& $\Gamma$&$Z_s$& $\frac{M_{Est}}{R_{Est}}$ \\
\hline
$L_{01}$&$Her~X-1$\cite{Rawls2011}& $2.1385$&$0.9283$&$0.2206$ \\
\hline
$L_{02}$&$4U~1538-52$\cite{Ozel2009}& $2.1434$&$0.9317$&$0.2602$\\
\hline
$L_{03}$&$SAX~J1808.4-3658$\cite{Demorest2010}& $2.1476$& $0.9483$&$0.2586$ \\
\hline
$L_{04}$&$LMC~X-4$\cite{Antoniadis2013}& $2.1509$&$0.9721$&$0.2601$\\
\hline
\rowcolor{gray!20}
$L_{05}$&$Cen~X-4$\cite{Maurya2023}& $2.1528$&$0.9802$&$0.2625$ \\ \hline
\rowcolor{gray!20}
$L_{06}$&$SMC~X-1$\cite{Maurya2024}& $2.1536$&$0.9925$&$0.2570$ \\ \hline 
\rowcolor{gray!20}
$L_{07}$&$GW170817-2$\cite{Maurya2024}& $2.1407$&$1.0238$&$0.2581$ \\ \hline 
$L_{08}$&$SMC~X-4$\cite{Cromartie2020}& $2.1552$&$1.0363$&$0.2766$ \\
\hline
$L_{09}$&$EXO~1785-248$\cite{Romani2022}& $2.1573$&$1.0542$&$0.2926$\\
\hline
\rowcolor{gray!20}
$L_{10}$&$PSR~J0030+0451$\cite{Maurya2025MassGap}& $2.1581$&$1.0551$&$0.2882$ \\ \hline
\rowcolor{gray!20}
$L_{11}$&$X7~(47~Tuc)$\cite{Bogdanov2016}& $2.1592$&$1.0568$&$0.3047$ \\ \hline
\rowcolor{gray!20}
$L_{12}$&$Swift~J1818.0-1607$\cite{Singh2026}& $2.1604$&$1.0574$&$0.3031$ \\ \hline
\rowcolor{gray!20}
$L_{13}$&$PSR~B0943+10$\cite{Maurya2026JCAP}& $2.1625$&$1.0582$&$0.3049$ \\ \hline
\rowcolor{gray!20}
$L_{14}$&$M13~neutron~star$\cite{Shaw2018}& $2.1648$&$1.0596$&$0.3032$ \\ \hline
$L_{15}$&$Cen~X-3$\cite{delaCruzDombriz2009}& $2.1665$&$1.0629$&$0.3347$ \\
\hline
\rowcolor{gray!20}
$L_{16}$&$PSR~B0656+14$\cite{Maurya2022JCAP}& $2.1682$&$1.0639$&$0.3099$ \\ \hline
\rowcolor{gray!20}
$L_{17}$&$XTE~J1739-285$\cite{Kaaret2007}& $2.1716$&$1.0662$&$0.3307$ \\ \hline
\rowcolor{gray!20}
$L_{18}$&$4U~1724-307$\cite{Hansraj2024}& $2.1734$&$1.0694$&$0.3151$ \\ \hline
$L_{19}$&$4U~1820-30$\cite{Kaaret1999}  & $2.1758$&$1.0731$&$0.3275$ \\
\hline
\rowcolor{gray!20}
$L_{20}$&$Aql~X-1$\cite{Li2017}& $2.1761$&$1.0750$&$0.3398$ \\ \hline
\rowcolor{gray!20}
$L_{21}$&$KS~1731-260$\cite{Maurya2025FQT}& $2.1770$&$1.0768$&$0.3493$ \\ \hline
$L_{22}$&$PSR~J1903+327$\cite{Linares2018}& $2.1774$&$1.0794$&$0.3140$ \\
\hline
\rowcolor{gray!20}
$L_{23}$&$Cyg~X-2$\cite{Casares2010}& $2.1809$&$1.0805$&$0.3191$ \\ \hline
$L_{24}$&$4U~1608-52$\cite{Freire2011}& $2.1839$&$1.0821$&$0.3267$ \\
\hline
$L_{25}$&$Vela~X-1$\cite{Abubekerov2008}& $2.1885$&$1.0847$&$0.3174$ \\
\hline
$L_{26}$& $4U~1724-207$\cite{Guver2010a}      & $2.1918$&$1.0864$&$0.3088$\\
\hline
$L_{27}$ & $SAX~J1748.9-2021$\cite{Guver2010b} & $2.1930$&$1.0935$&$0.3366$ \\
\hline
\rowcolor{gray!20}
$L_{28}$&$4U~1702-429$\cite{Singh2022Lovelock}& $2.1957$&$1.0972$&$0.3319$ \\ \hline
$L_{29}$ & $PSR~J1614-2230$ \cite{Elebert2009}  & $2.2034$&$1.1130$&$0.3121$\\
\hline
$L_{30}$&  $PSR~J0348+0432$\cite{Abubekerov2008}     & $2.2117$&$1.1209$& $0.3037$\\
\hline
$L_{31}$&  $PSR~J0740+6620$\cite{Linares2018}     & $2.2308$&$1.1288$& $0.3297$\\
\hline
$L_{32}$&  $PSR~J1748-2446ad$\cite{Elebert2009} & $2.2374$&$1.1319$& $0.3467$ \\
\hline
$L_{33}$ & $PSR~J2215+5135$ \cite{delaCruzDombriz2009}      & $2.4391$&$1.1361$& $0.3292$\\
\hline
$L_{34}$ & $PSR~J0952-0607$\cite{Rawls2011}   & $2.4407$&$1.1382$& $0.3986$ \\
\hline
\rowcolor{gray!20}
$L_{35}$&$4U~1700-37$\cite{Hansraj2025}& $2.4436$&$0.1391$&$0.3703$ \\ \hline
\rowcolor{gray!20}
$L_{36}$&$GW190814-2$\cite{Maurya2024ApJ}& $2.4529$&$0.1407$&$0.3954$ \\ \hline
\end{tabular}
\end{center}
\vspace{2mm}
\footnotesize
\begin{justify}
\textit{\bf{Table~6:}} presents the numerical values of the radial adiabatic index ($\Gamma$), surface redshift ($Z_s$), and compactness parameter corresponding to the maximum estimated mass configuration, expressed through the ratio of the maximum estimated mass ($M_{\rm Est}$) to the associated stellar radius ($R_{\rm Est}$) at which the maximum mass is attained. An examination of the tabulated results reveals that the surface redshift satisfies the condition $Z_s \leq 5$ for all the compact-star configurations considered in the present study, which is consistent with the accepted upper bound for stable anisotropic stellar systems \cite{Bowers1974,Ivanov2002}. Furthermore, the values of the radial adiabatic index satisfy the stability criterion $\Gamma > 4/3$ throughout the considered sample, thereby confirming the dynamical stability of the corresponding stellar configurations against infinitesimal radial perturbations \cite{Chandrasekhar1964}. In addition, the compactness parameter remains within the Buchdahl limit, $u=M_{\rm Est}/R_{\rm Est}<4/9$, ensuring that all the stellar models satisfy the fundamental compactness constraint imposed by GR \cite{Buchdahl1959}. Therefore, the numerical results obtained for the 36 compact stars collectively demonstrate that the proposed stellar model is consistent with the established theoretical requirements for realistic compact star model.
\end{justify}
\end{table}
\begin{figure}[H]
\begin{center}
\includegraphics[keepaspectratio=true,height=4in, width=7.5in]{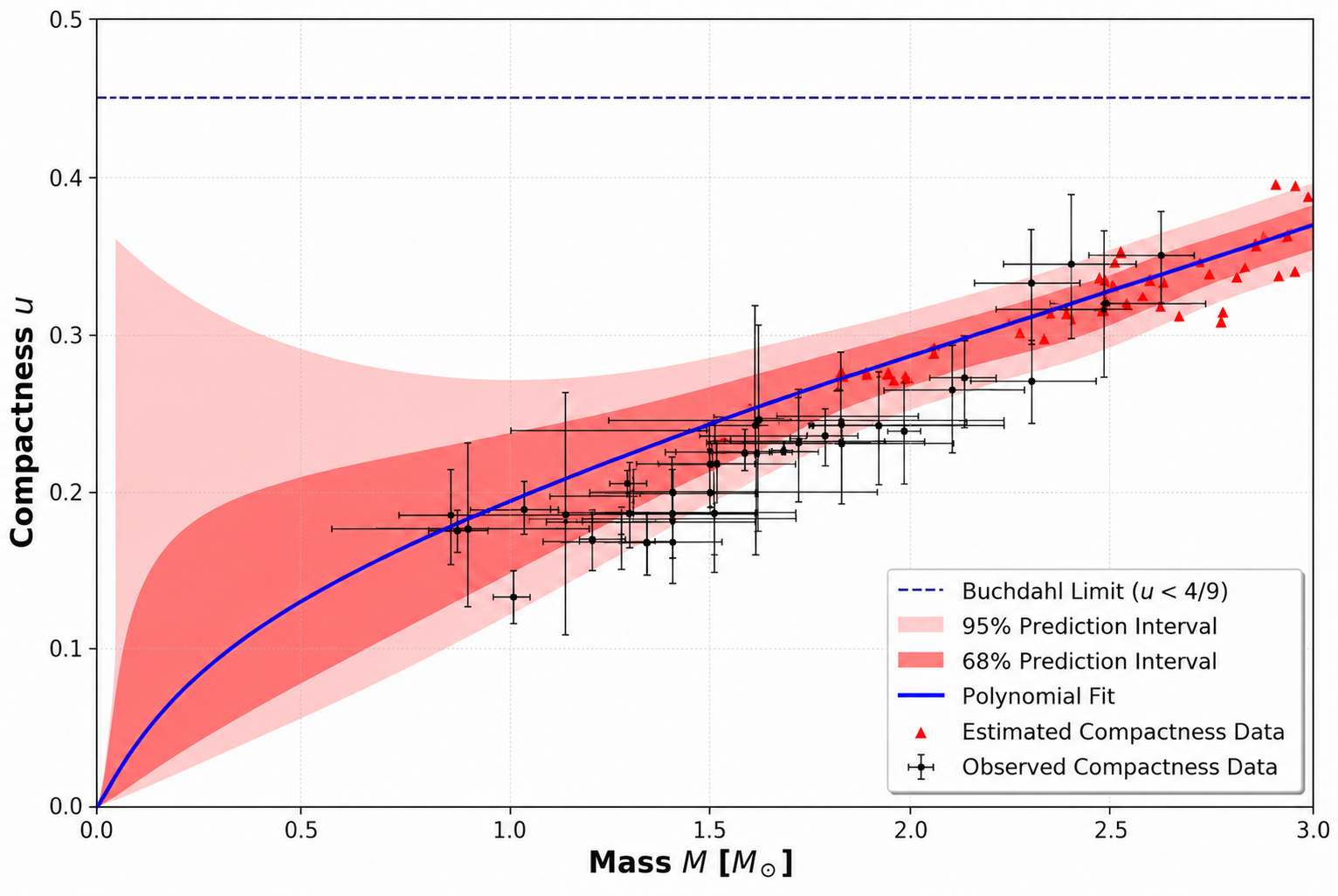}\\
\begin{justify}
Figure~12. Comparison of the compactness profile of neutron stars obtained from observational constraints and the theoretical model. The black circles with error bars correspond to observationally inferred compactness values, whereas the red triangles represent the compactness predicted by the present theoretical model. A cubic polynomial regression is performed on the theoretical data to obtain the best-fit relation (solid blue line). The associated $68\%$ and $95\%$ prediction intervals are shown by the dark and light red shaded bands, respectively. The dashed horizontal line indicates the Buchdahl limit line ($u = 4/9$), which serves as the theoretical upper bound for the compactness of a static isotropic relativistic star. The close agreement between the observational compactness measurements and the theoretical predictions demonstrates that the proposed stellar model provides a statistically consistent description of the observed mass--compactness relation for neutron stars.\\
\end{justify}
\end{center}
\end{figure}
\noindent\begin{itemize}

\item \begin{itemize}

\item Figure~12 presents a comparison between the compactness inferred from observational measurements and the compactness predicted by the proposed stellar model for the neutron-star sample considered in the present work. The observational compactness is evaluated from the observed mass ($M_{\mathrm{Obs}}$) and the reported radius ($R_{\mathrm{Rep}}$) listed in Table~\ref{tab:maxmassa} using the relation
\[
u=\frac{M}{R},
\]
and is represented by the black data points together with their associated uncertainties. The red triangular markers denote the compactness values predicted by the proposed stellar model, whereas the solid blue curve represents the cubic polynomial regression obtained from these theoretical predictions. The corresponding $68\%$ and $95\%$ prediction intervals are illustrated by the dark and light red shaded regions, respectively, while the horizontal dashed line indicates the Buchdahl compactness limit ($u<4/9$).

~~~It is evident from the figure that the compactness increases monotonically with stellar mass, reflecting the progressive gravitational compactification of neutron stars. Lower-mass configurations ($\sim0.8$--$1.2\,M_{\odot}$) exhibit comparatively smaller compactness, whereas more massive neutron stars possess systematically larger compactness, consistent with the stronger gravitational fields expected in compact stellar configurations.

~~~An important outcome of Figure~12 is the excellent agreement between the observationally inferred compactness and the theoretical predictions. The polynomial regression successfully captures the overall trend of the theoretical compactness values, while the overwhelming majority of the observational measurements are enclosed within the corresponding $95\%$ prediction interval. The comparatively narrower prediction bands over the intermediate and high-mass regions indicate that the theoretical predictions remain highly consistent where the compactness data are more densely distributed. In contrast, the wider prediction interval toward the lower-mass end reflects the comparatively sparse sampling of theoretical data in that region and the corresponding increase in predictive uncertainty.

~~~Another significant result is that both the observational compactness values and the theoretical predictions remain comfortably below the Buchdahl limit,
\[
u<\frac49,
\]
throughout the entire mass range considered. This demonstrates that all the stellar configurations obtained from the proposed model satisfy the fundamental compactness criterion required for physically acceptable relativistic stars. Moreover, the smooth monotonic increase of compactness without any abrupt deviation indicates that the model provides a stable and self-consistent description of the mass--compactness relation.

~~~The present results are also consistent with recent observational and theoretical studies. In particular, NICER measurements and recent mass--radius analyses infer neutron-star radii of approximately $10$--$14\,\mathrm{km}$, leading to compactness values comparable to those obtained in the present work \cite{Miller2019,Riley2019}. Likewise, realistic relativistic stellar models predict a similar monotonic increase of compactness with stellar mass while remaining below the Buchdahl limit \cite{Lattimer2016,Lattimer2007,Ozel2016}. The agreement obtained here therefore provides an independent validation of the proposed stellar model against current observational constraints.

~~~Overall, Figure~12 demonstrates that the compactness predicted by the proposed stellar model is statistically consistent with the observationally inferred compactness over the entire mass range considered. The close agreement between the observational measurements, the theoretical predictions, the polynomial regression, and the associated prediction intervals, together with the fulfillment of the Buchdahl compactness bound, further supports the mathematical consistency, physical acceptability, and observational viability of the proposed anisotropic stellar model.

\end{itemize}
\end{itemize}
\begin{figure}[H]
\begin{center}
\fbox{
\includegraphics[keepaspectratio=true,height=4in, width=7.5in]{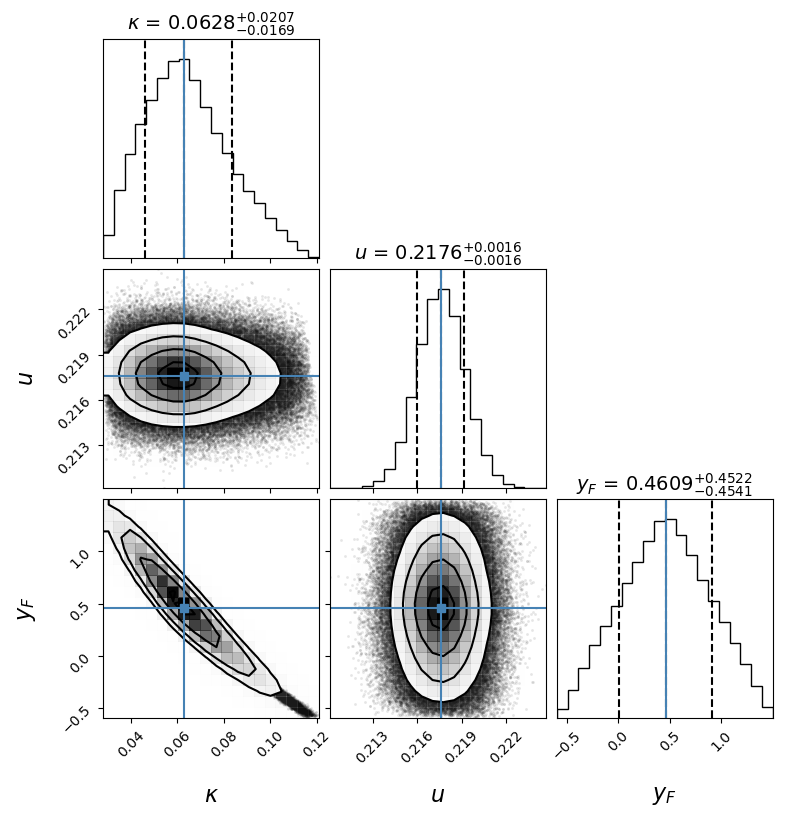}}\\
\end{center}
\begin{justify}
Figure~13: The compactness parameter is defined as $u_{\mathrm{obs}}=\frac{M_{\mathrm{obs}}\,M_{\odot}}{R_{\mathrm{obs}}}$, and its associated uncertainty is given by $\sigma_u=\sqrt{\left(\frac{\sigma_M}{M_{\mathrm{obs}}}\right)^2+\left(\frac{\sigma_R}{R_{\mathrm{obs}}}\right)^2}$. The fitting function is assumed to follow a linear relation of the form $y_F=a-bu$, where $a$ and $b$ are free parameters. The parameter $\kappa$ is obtained from equation~\eqref{arpann}. The priors are chosen as $0.05<u<0.35$, $0.5<a<1.5$, and $0.0<b<5.0$. Parameter estimation is performed using a chi-square minimization approach, with sampling carried out via the Markov Chain Monte Carlo (MCMC) method. The best-fit values correspond to the median (50th percentile) of the posterior distributions, while uncertainties are determined from the $1\sigma$ (68\%) credible intervals. Contour levels corresponding to $1\sigma$, $2\sigma$, and $3\sigma$ confidence regions are also shown. The acceptance fraction was approximately $0.5384^{+0.0090}_{-0.0077}$. The MCMC chains satisfied the convergence criterion ($\hat{R}<1.01$).\\
\end{justify}
\end{figure}
\noindent
\begin{itemize}

\item Figure~13 presents the posterior probability distributions of the three free parameters entering the tidal-deformability analysis, namely the tidal Love number $\kappa$, the compactness parameter $u$, and the perturbation parameter $y_F$. These posterior distributions are obtained by fitting equation~\eqref{arpann} to the observational mass--radius data listed in Table~\ref{tab:maxmassa}. The statistical methodology, including the adopted priors, likelihood function, acceptance fraction, and convergence criterion, is summarized in the figure caption. The diagonal panels display the marginalized one-dimensional posterior distributions of the individual parameters, whereas the off-diagonal panels show the corresponding joint posterior distributions together with the $1\sigma$, $2\sigma$, and $3\sigma$ confidence regions.

~~~The MCMC analysis yields the following median parameter estimates:
\[
\kappa = 0.0628^{+0.0207}_{-0.0169},\qquad
u = 0.2176^{+0.0016}_{-0.0016},\qquad
y_F = 0.4609^{+0.4522}_{-0.4541},
\]
where the quoted uncertainties correspond to the $68\%$ credible intervals. Among the three parameters, the compactness exhibits the narrowest posterior distribution, indicating that it is the most tightly constrained quantity by the observational mass--radius data. In contrast, the posterior distributions of $\kappa$ and, particularly, $y_F$ are comparatively broader, reflecting the larger uncertainties associated with the tidal response and the internal metric perturbation of the stellar configuration.

~~~The joint posterior distributions provide important insight into the correlation structure of the model parameters. A pronounced negative correlation is observed between $\kappa$ and $y_F$, indicating that different combinations of these parameters can produce nearly identical tidal responses, thereby preserving agreement with the observational constraints. The $\kappa$--$u$ plane exhibits only a weak correlation, whereas the $u$--$y_F$ plane indicates a comparatively weak dependence, suggesting that the compactness is primarily determined by the observational mass--radius measurements rather than by the internal perturbation parameter.

~~~An important physical implication of this analysis is that the observational data restrict the allowed parameter space of the proposed stellar model to a unique and well-defined region. The absence of secondary maxima or disconnected confidence regions indicates that the adopted model possesses a stable and statistically consistent solution within the explored parameter space. Consequently, the compactness inferred from the MCMC analysis remains fully consistent with the mass--compactness relation discussed in Figure~12, while the inferred tidal Love number supports the tidal-deformability analysis presented in Figure~6.

~~~The statistical reliability of the parameter estimation is further supported by the MCMC diagnostics. The acceptance fraction,
\[
0.5384^{+0.0090}_{-0.0077},
\]
lies within the range generally regarded as suitable for efficient sampling, and the convergence criterion ($R<1.01$) confirms that the Markov chains have converged satisfactorily. These diagnostics indicate that the inferred posterior distributions are statistically robust and are not influenced by inadequate sampling or poor convergence.

~~~Overall, Figure~13 provides an independent statistical validation of the proposed compact-star model. The close agreement between the inferred posterior distributions and the observational mass--radius data demonstrates that the model simultaneously reproduces the observed compactness, remains consistent with the tidal-deformability analysis, and yields physically acceptable estimates of the principal stellar parameters. Consequently, the MCMC analysis further strengthens the mathematical consistency, physical acceptability, and observational viability of the proposed anisotropic stellar model within the framework of Einstein's gravity.
\end{itemize}
\section{Brief Discussions and Conclusion}

In the present work, we have developed an anisotropic compact-star model within the framework of Einstein's General Relativity by employing the Buchdahl metric-I ansatz together with a generalized polynomial equation of state. The resulting interior solution was matched smoothly to the exterior Schwarzschild space-time through the standard Darmois--Israel junction conditions, leading to a class of physically acceptable stellar configurations capable of describing realistic neutron stars over a broad range of observed masses.

A comprehensive physical analysis demonstrates that the proposed model satisfies all the essential requirements expected for realistic relativistic stellar configurations. The energy density and both radial and tangential pressure components remain finite, positive, and monotonically decreasing throughout the stellar interior, while the anisotropic factor vanishes at the stellar centre and increases smoothly toward the surface, ensuring the regularity of the interior solution. The generalized Tolman--Oppenheimer--Volkoff equilibrium equation is satisfied throughout the stellar interior through the combined action of the gravitational, hydrostatic, and anisotropic forces. The revised force analysis confirms that these three forces remain in mutual balance at every interior point, thereby establishing complete hydrostatic equilibrium for the obtained stellar configurations.

The proposed model also satisfies the principal stability and causality criteria. The adiabatic index remains greater than the Chandrasekhar stability limit ($\Gamma>4/3$), confirming stability against infinitesimal radial perturbations. Furthermore, the radial and tangential sound speeds remain subluminal throughout the stellar interior and satisfy the Herrera cracking criterion, indicating that the configurations are dynamically stable and free from local anisotropic instabilities. The calculated compactness remains below the Buchdahl limit, while the corresponding surface redshift lies well within the theoretically accepted range for anisotropic compact stars, confirming the physical acceptability of the obtained solutions.

The observational relevance of the proposed model has been examined through a detailed comparison with a large sample of observed neutron stars. The predicted mass--radius configurations are found to be consistent with the available observational constraints, and the corresponding stellar parameters remain within physically acceptable ranges. The estimated maximum masses obtained from the present model are systematically higher than the corresponding observed masses, indicating that positive pressure anisotropy provides additional support against gravitational collapse while preserving equilibrium and stability. Interestingly, several of the estimated maximum masses fall within the lower mass-gap region, suggesting the possibility that stable neutron stars may exist in this domain. Although this observation requires future confirmation through improved astronomical measurements, it highlights an interesting astrophysical implication of the present anisotropic framework.

The tidal properties of the constructed stellar model are also found to be compatible with current multimessenger observations. The tidal deformability decreases monotonically with increasing stellar mass and remains consistent with the observational constraints inferred from GW170817 and recent NICER measurements. In addition, the compactness analysis demonstrates excellent agreement between the theoretical predictions and the observationally inferred compactness over the entire mass range considered. The polynomial regression and the associated prediction intervals further establish the statistical consistency of the proposed mass--compactness relation with the available observational data.

An additional strength of the present work is provided by the Bayesian parameter estimation. The Markov Chain Monte Carlo analysis yields well-constrained posterior distributions for the tidal Love number, compactness parameter, and the perturbation parameter. The inferred posterior distributions exhibit stable convergence and reveal a unique physically admissible region of the parameter space capable of reproducing the observed mass--radius relation. The statistical consistency demonstrated by the compactness reconstruction and the Bayesian analysis provides an independent validation of the proposed stellar model beyond the conventional physical diagnostics.

Taken together, the analytical solutions, graphical investigations, numerical estimates, observational comparisons, compactness reconstruction, and Bayesian parameter estimation consistently support the mathematical consistency, physical acceptability, and observational viability of the proposed anisotropic stellar model. The present study demonstrates that the inclusion of anisotropic stresses within Einstein's gravity provides a physically consistent mechanism for constructing stable neutron-star configurations that simultaneously satisfy the requirements of regularity, equilibrium, causality, dynamical stability, compactness, and tidal deformability while remaining compatible with current astrophysical observations.

The present investigation may be extended in several directions. Future studies may incorporate rotation, strong magnetic fields, more realistic microscopic equations of state, or alternative theories of gravity to investigate the behavior of ultra-dense matter under more general astrophysical conditions. Such extensions may further clarify the role of anisotropy in determining the structure, evolution, and observable properties of compact stars.

\section*{Acknowledgments}
RB gratefully acknowledges the Inter-University Centre for Astronomy and Astrophysics (IUCAA), Pune, India, for the Visiting Associateship. AB gratefully acknowledges the financial support received through the UGC-NET Senior Research Fellowship (SRF). SP gratefully acknowledges the financial support received through the Swami Vivekananda Merit-cum-Means Scholarship for Ph.D. Research Fellowship, Higher Education Department, Government of West Bengal.
\section*{Funding}
There is no funding to report for this article.
\section*{Data Availability Statement}
This article has no associated data as the work is completely theoretical in nature.
\section*{Code Availability Statement}
My manuscript has no associated code/software.
\section*{Conflict of Interest}
The authors declare that they have no known competing financial interests or personal relationships that could have appeared to influence the work reported in this paper.

\bibliographystyle{ieeetr}
\bibliography{references}

\end{document}